\documentclass[aps,pra,twocolumn,showpacs,superscriptaddress,groupedaddress,amsmath,amssymb]{revtex4-1}

\usepackage{subfigmat}

\usepackage{xparse} 
\usepackage{graphicx}
\usepackage{subfigure}
\usepackage{amsmath}
\usepackage{color}
\usepackage{longtable}
\usepackage{dsfont}
\usepackage{hyperref}
\usepackage{cleveref}
\usepackage[margin=50pt]{geometry}

\newcommand{\braket}[1]{\ensuremath{\langle#1\rangle}}
\newcommand{\bs}[1]{\ensuremath{\boldsymbol{#1}}}
\newcommand{\md}[1]{\ensuremath{\mathds{#1}}}
\newcommand{\mbb}[1]{\ensuremath{\mathbb{#1}}}
\newcommand{\ket}[1]{\ensuremath{|#1\rangle}}
\newcommand{\bra}[1]{\ensuremath{\langle #1 |}}

\DeclareMathOperator{\Tr}{Tr}

\DeclareDocumentCommand\op { m g } {%
	\IfNoValueTF {#2} {%
	\ensuremath{|#1\rangle\!\langle #1|}
	}{%
	\ensuremath{|#1\rangle\!\langle #2|}
	}%
}
\DeclareDocumentCommand\ip { m g } {%
	\IfNoValueTF {#2} {%
		\ensuremath{\langle #1|#1\rangle}
	}{%
		\ensuremath{\langle #1|#2\rangle}
	}%
}

\begin{document}

\title{Combining $T_1$ and $T_2$ estimation with randomized benchmarking and bounding the diamond distance}
\author{Hillary Dawkins$^{1,2}$, Joel Wallman$^{1,3}$, and Joseph Emerson$^{1,3,4}$}
\affiliation{
Institute for Quantum Computing$^{1}$, the Department of Physics and Astronomy$^2$, and the Department of Applied Mathematics$^3$,\\
University of Waterloo, Waterloo, Ontario N2L 3G1, Canada\\
Canadian Institute for Advanced Research$^4$, Toronto, Ontario M5G 1Z8, Canada
}
\date{\today}

\begin{abstract}
The characterization of errors in a quantum system is a fundamental step for two important goals. First, learning about specific sources of error is essential for optimizing experimental design and error correction methods. Second, verifying that the error is below some threshold value is required to meet the criteria of threshold theorems. We consider the case where errors are dominated by the generalized damping channel (encompassing the common intrinsic processes of amplitude damping and dephasing) but may also contain additional unknown error sources. 
We demonstrate the robustness of standard $T_1$ and $T_2$ estimation methods and provide expressions for the expected error in these estimates under the additional error sources. 
We then derive expressions that allow a comparison of the actual and expected results of fine-grained randomized benchmarking experiments based on the damping parameters.
Given the results of this comparison, we provide bounds that allow robust estimation of the thresholds for fault-tolerance.   
\end{abstract}

\maketitle

\section{Introduction} 

Achieving practical quantum computation requires quantum systems to be precisely controlled despite being prone to intrinsic errors. Characterizing the errors acting on a system is of the utmost importance in the quest to improve control. 
There are two paradigmatic approaches to error characterization for two-level quantum systems. In the first approach, one assumes a system-specific  but control-independent error model and then estimates the corresponding parameters of the model, with the most common protocols here being damping parameter measurements which characterize fundamental Markovian noise processes such as amplitude damping (from spontaneous absorption and emission) and dephasing errors (associated with inhomogeneous fields) \cite{NC:2005, Ladd:2010}, commonly referred to as $T_1$ and $T_2$ measurements.
In the second approach, one does not make any assumptions about the error model (beyond the usual Markovian assumption) but directly estimates the overall effective error rates. The overall error-rate would include both the control errors (where control is required to generate quantum gates) and the control-independent noise affecting the system. The most common approach here is randomized benchmarking experiments that uniformly sample from the Clifford group \cite{Emerson:2005, Dankert:2006, Levi:2007, Magesan:2011, Magesan:2012, Wallman:2017}. 

In this paper we show how the information from both approaches may be combined to establish a tight connection between these standard experimental figures of merit and the error threshold condition in fault-tolerant threshold theorems \cite{Shor:1995, Devitt:2013}. In general one can not simply compare the error rate from a randomized benchmarking experiment directly to fault-tolerant thresholds  unless the error model and the threshold theorem satisfy a few critical assumptions. Fault-tolerant threshold theorems typically rely on a norm-based measure of gate error, with the diamond distance \cite{Kitaev:1997} between the error channel and the identity channel being by far the most common metric. However norm-based measures such as the diamond distance are not directly measurable through any of the above experiments \cite{Sanders:2015}.  

Below we develop a general framework for error characterization that compares the estimates of standard damping parameters and randomized benchmarking experiments. This enables an important test of the relative role of experimental control errors by comparing the experimentally obtained randomized benchmarking error rates with  predicted values inferred from the damping parameters.  We then show how any measured discrepancies give rise to a robust bound on estimates of the diamond distance. We consider cases where the error model is close enough to the generalized damping channel, which includes both amplitude damping and dephasing processes at finite temperature, that deviations from this model can be characterized perturbatively. If the measured error rate and unitarity are close enough to their predicted values, this bound provides a significant improvement over more general bounds obtained previously \cite{Wallman:2015Nov}.

This paper is organized as follows. 
First, we consider how amplitude damping and dephasing estimation can be affected by additional, unknown, intrinsic errors.
We show that amplitude damping estimation is mostly insensitive to these additional noise sources, while dephasing estimation can be affected greatly when measured via a static Ramsey experiment. We propose an alternative method for Ramsey experiments that proves insensitive to these additional noise sources, and provide a numerical demonstration of its effectiveness. Second, we show how the measured parameters can be used to predict the randomized benchmarking error rate and unitarity when the error model is given by an ideal generalized damping channel. This enables an important test of the relative role of experimental control errors by comparing the experimentally obtained randomized benchmarking error rates with the predicted values.  
Finally, we derive upper bounds on the diamond distance (between the generalized damping channel and the identity map), and provide an extension to this bound that accounts for the additional perturbative errors as estimated by the randomized benchmarking experiments.        

\section{Ideal Generalized Damping Model}

Experimental two-level systems typically experience both amplitude and phase damping. Amplitude 
damping is caused by relaxations to, and excitations from, 
the ground state at some net rate $\Gamma_1$, leading to an equilibrium 
population $\lambda$ in the ground state ($\lambda = 1$ at zero temperature). 
Phase damping is caused by weak coupling to a bath that causes 
coherences between the energy eigenstates to decay at a rate $\Gamma_2$. These 
processes can be described by a generalized damping master equation
\begin{align}
\frac{\partial \rho}{\partial t} = \lambda\Gamma_1G[\rho,\sigma_+] + 
(1-\lambda)\Gamma_1G[\rho,\sigma_-] + \frac{\Gamma_2}{2}G[\rho,Z],
\label{eqn:GD_ME}
\end{align}
where $G[\rho,M,N] = M\rho N^\dagger - (N^\dagger M\rho + \rho N^\dagger M)/2$ 
and $G[\rho,M]=G[\rho,M,M]$. The excitation operators are defined as $\sigma_+ = 1/\sqrt{2}(X+iY)$, $\sigma_- = 1/\sqrt{2}(X - iY)$, and $X$, $Y$, $Z$ are the 2-qubit Pauli operators. 
Note that applying corrective pulse sequences such as Hahn-echo \cite{Hahn:1950} or Carr-Purcell-Meiboom-Gill
\cite{Carr:1954, Meiboom:1958} gives the same generalized 
damping model with a redefinition of the meaning of $\Gamma_2$ (i.e. the model describes the effective rate of dephasing, whether correction pulses have been applied or not).

\section{Perturbed Generalized Damping Model}

Generalized damping is believed to account for the majority of the intrinsic
noise in experimental systems, as shown by the prominence of $\Gamma_1$ and $\Gamma_2$
values. However, it is likely that there are other unknown intrinsic noise sources, which
can be viewed as perturbations from a generalized damping process as follows.
The most general form of a master equation for a single qubit is
\begin{align}\label{eq:master_equation}
	\frac{\partial \rho}{\partial t} 
	= -i[H,\rho] + \sum_{i,j = 2}^{4}A_{i,j}G[\rho,B_i,B_j]
\end{align}
where the Hamiltonian is
\begin{align}
H = \frac{1}{2i}(F^\dagger - F), ~~ F = \frac{1}{\sqrt{2}}\sum_{i=2}^4 A_{1,i}B_i,
\end{align}
$B = \{I/\sqrt{2},\sigma_+,\sigma_-,Z/\sqrt{2}\}$, and $A$ is a coefficient 
matrix. 
The first term of \cref{eq:master_equation} describes the Hamiltonian dynamics of the system, while the second part describes the noisy dissipator terms. 
The Hamiltonian can be estimated separately \cite{Schirmer:2004, deClercq:2016}, and so we assume that the Hamiltonian is known and does not contribute to the error model.  
Therefore we set $H=0$ and so delete the first row and column of $A$. 
The unknown intrinsic errors are described by the remaining submatrix $A_{i,j \in \{2,3,4\}} $ (hereafter denoted simply as $A$).
The remaining matrix must be Hermitian and positive semi-definite for the evolved state to
be a valid quantum state \cite{Breuer}. 
A general master equation with coefficient matrix
\begin{align}\label{eq:general_A}
A = \left(
\begin{array}{ccc}
 \lambda  \Gamma_1 & \alpha_r-i \alpha_i & \beta^*  \\
 i \alpha_i+\alpha_r & (1-\lambda ) \Gamma_1 & \delta^*  \\
 \beta  & \delta  & \Gamma_2 \\
\end{array}
\right)
\end{align} 
is then a perturbation about a generalized damping channel if the off-diagonal
elements of $A$ are small compared to the diagonal elements. We can also assume
all variables in $A$ are real without loss of generality by including any phase
from $\beta$ and $\delta$ into $\sigma_{\pm}$ respectively.
In general, the dynamics described by $A$ will be a more complicated damping process (i.e. the off-diagonal interaction terms will break the symmetry of phase damping and couple the $\Gamma_1$ and $\Gamma_2$ processes). 

For a general state $\rho =(I + \bs{r}\cdot\bs{\sigma})/2$, where $\bs{r}$ is 
the Bloch vector and $\bs{\sigma} = (X,Y,Z)$, substituting \cref{eq:general_A}
into \cref{eq:master_equation} gives
\begin{align}\label{eq:bloch_PDE}
\frac{\partial \bs{r}}{\partial t} 
&= C\bs{r} + \bs{\lambda}
\end{align}
where 
\begin{align}\label{eqn:C}
C &= \left(
\begin{array}{ccc}
 \alpha_r-\Gamma_2' & \alpha_i & \beta  \\
 \alpha_i & -\alpha_r-\Gamma_2' & 0 \\
 \beta  & 0 & -\Gamma_1 \\
\end{array}
\right), \notag\\
\bs{\lambda} &= \left(\begin{array}{c}
2 \sqrt{2} \delta -2 \beta \\ 0 \\ \Gamma_1 (2 \lambda -1)
\end{array}\right) ,
\end{align}
the total dephasing rate is $\Gamma_2' = \frac{\Gamma_1}{2}+\Gamma_2$, and we have redefined
$\frac{\beta +\delta }{\sqrt{2}}\to\beta$. 
Also note that \cref{eqn:C} describes the evolution of a Bloch sphere where the $x$ and $y$ axes have been redefined by a rotation around the $z$ axis by $\phi = (\text{arg}(\beta) - \text{arg}(\delta))/2$ due to the phase absorption of $\beta$ and $\delta$ into the basis of \cref{eq:general_A}. This redefinition does not affect later arguments of the robustness of parameter estimation.
As $C$ is Hermitian, it has an
orthonormal set of eigenvectors $V = (v_1,v_2,v_3)$ with associated eigenvalues 
$\bs{\eta} = (\eta_1,\eta_2,\eta_3)$. Therefore, under a generic master equation, an initial state 
$\rho$ with Bloch vector $\bs{r}(0) = \sum_j c_j(0) v_j$ evolves to the state with 
Bloch vector $\bs{r}(t) = \sum_j c_j(t) v_j$ at time $t$ where
\begin{align}
\frac{\partial c_j(t)}{\partial t} = \eta_j c_j(t) + \lambda_j,
\end{align}
and $\lambda_j = \braket{v_j|\lambda}$, which can be exactly solved to give
\begin{align}\label{eq:evolved_bloch}
c_j(t) = \frac{\lambda_j(e^{\eta_j t}-1)}{\eta_j} + c(0)e^{\eta_j t}.
\end{align}
The three different eigenvalues result in three characteristic decay time 
scales. Also note that the dephasing process is no longer around the ground state of the system, since $\bs{\lambda}$ is not an eigenvector of $C$ in general.

The remaining difficulty in determining the evolution of a generic state is to 
identify the eigenvalues and eigenvectors.
As a simple example, when $\alpha_i = \beta = \delta = 0$ (more general than normally 
considered as $\alpha_r$ can be nonzero), $V = \md{I}_3$ and so a state with 
Bloch vector $\bs{r} = (r_x,r_y,r_z)$ evolves to a state with Bloch vector 
components
\begin{align}\label{eq:GDexample}
r_x(t) &= r_x e^{t \left(\alpha_r-\Gamma_2'\right)}, \notag\\
r_y(t) &= r_y e^{t \left(-\left(\Gamma_2'+\alpha_r\right)\right)}, \notag\\
r_z(t) &= r_z e^{-\Gamma_1 t}-(2 \lambda -1) \left(e^{-\Gamma_1 t}-1\right).
\end{align}
Any non-zero value of $\alpha_r$ splits the degeneracy of the first two 
eigenvalues so that there is no longer a single dephasing lifetime. This is 
reflected in the sensitivity of $\Gamma_2'$ experiments to perturbations as 
discussed in more detail later.  

Any experiment to estimate the generalized damping parameters will include state preparation and measurement (SPAM) 
errors. Applying a (perfect) random operation that leaves the ideal preparation 
$\psi$ and measurement $M$ invariant immediately after preparing a state and 
prior to a measurement respectively reduces any SPAM to
\begin{align}\label{eq:SPAM}
\psi&\to \rho=\tfrac{1}{2}(I_2 + (1-k)\vec{r}\cdot\vec{\sigma}) \notag\\
M&\to M' =(1-n_1)M + n_2 I_2,
\end{align}
for some (small) positive constants $k$ and $n_1$ and (small) constant $n_2$. 
As the general master equation is trace-preserving and the trace is linear,
\begin{align}\label{eq:include_SPAM}
\Tr M'\rho(t) &= (1-n_1) \Tr M \rho(t)  + n_2.
\end{align} 
Moreover, by \cref{eq:evolved_bloch} the SPAM error simply reduces the 
coefficients of the exponentials and does not change the exponential decay 
rate. We set $k,n_1,n_2\to 0$ to simplify analytic expressions as the 
contributions are relatively small and the full expressions can be easily 
obtained using \cref{eq:include_SPAM}, but numerical simulations include 
non-zero values.

\section{Amplitude damping rate ($\Gamma_1$) estimation}

In a population inversion experiment, the system is prepared in the excited 
state $\op{1}$ $(r_z(0)=-1)$, then allowed to evolve for a time $t$ before measuring the 
relative population levels of $\op{0}$ and $\op{1}$ to obtain $Q_1(t) = 
\Tr(-Z\rho(t)) = -r_z(t)$. This preparation and measurement captures the rate of decay of the $-z$ axis of the Bloch sphere, or $-e_3$, where $\{e_k\}$ is the canonical basis of $\mbb{R}^3$. Under generalized damping, 
\begin{align}\label{eq:T1decay}
Q_1(t) &= 2\lambda e^{-\Gamma_1t}-2\lambda+1 = c_1e^{-\Gamma_1t} + c_0
\end{align}
by \cref{eq:GDexample} with $\bs{r}(0) = (0,0,-1)$. Therefore $\Gamma_1$ can be estimated by measuring $Q_1(t)$ for multiple values of $t$ and 
fitting to \cref{eq:T1decay}. 

In the perturbed case, measuring the rate of decay of $-e_3$ will essentially measure the rate of decay of the third eigenvector of $C$, $v_3$ (where the decay rate is given by the eigenvalue $\eta_3$). 
This can be seen by first letting $\alpha_i = 0$, for which we have $C\tilde{v}_3 = \tilde{\eta}_3 \tilde{v}_3$ where
\begin{align}
\tilde{v}_3^{T} &= \left(\sin\theta,0,-\cos\theta\right), \notag\\
\tan\theta &= \frac{g-\sqrt{4\beta^2+g^2}}{2\beta}, \notag \\
\tilde{\eta}_3 &= \frac{1}{2} \left(-g+\sqrt{4 \beta ^2+g^2}\right) - \Gamma_1, \notag \\
g &=\Gamma_2' - \alpha_r - \Gamma_1,
\end{align}
and $\tilde{}$ notation has been used to indicate the special case of $\alpha_i  = 0$.
For $\alpha_i\neq 0$,
\begin{align}
C\tilde{v}_3 - \tilde{\eta}_3 \tilde{v}_3 = \sin\theta\alpha_i e_2.
\end{align}
Therefore, in full generality, the third eigenvector is given by $v_3  = \tilde{v}_3 + O(\alpha_i\beta)$.
For perturbations about 
the generalized damping channel, $\beta\ll g$ implying $\tan\theta \in O(\beta)$, $\sin\theta \in O(\beta)$, and $\cos\theta \approx 1 + O(\beta^2)$ where we have also used $\Gamma_2' \gg \Gamma_1$.
The measurement essentially finds the rate of decay of $v_3$ since
\begin{align}
-e_3^\dagger v_3 &= \cos\theta + O(\alpha_i\beta) \notag \\
&= 1 + O(\beta^2) + O(\alpha_i\beta) \approx 1.
\end{align}
The measurement will pick up the decay of the other 2 eigenvectors on the order of $\beta^2$ and $\beta\alpha_i$, however these contributions are negligible compared to the first leading correction term on the estimate due to the perturbation of $\eta_3$ given by
\begin{align}
\eta_3 \approx -\Gamma_1 + \frac{\beta^2}{\Gamma_2' - \Gamma_1}
\end{align} 
(see Appendix D for full expressions of the eigenvalues of $C$). 

\begin{figure*}[h]
\centering
\subfigure[]{
\includegraphics[scale=0.6]{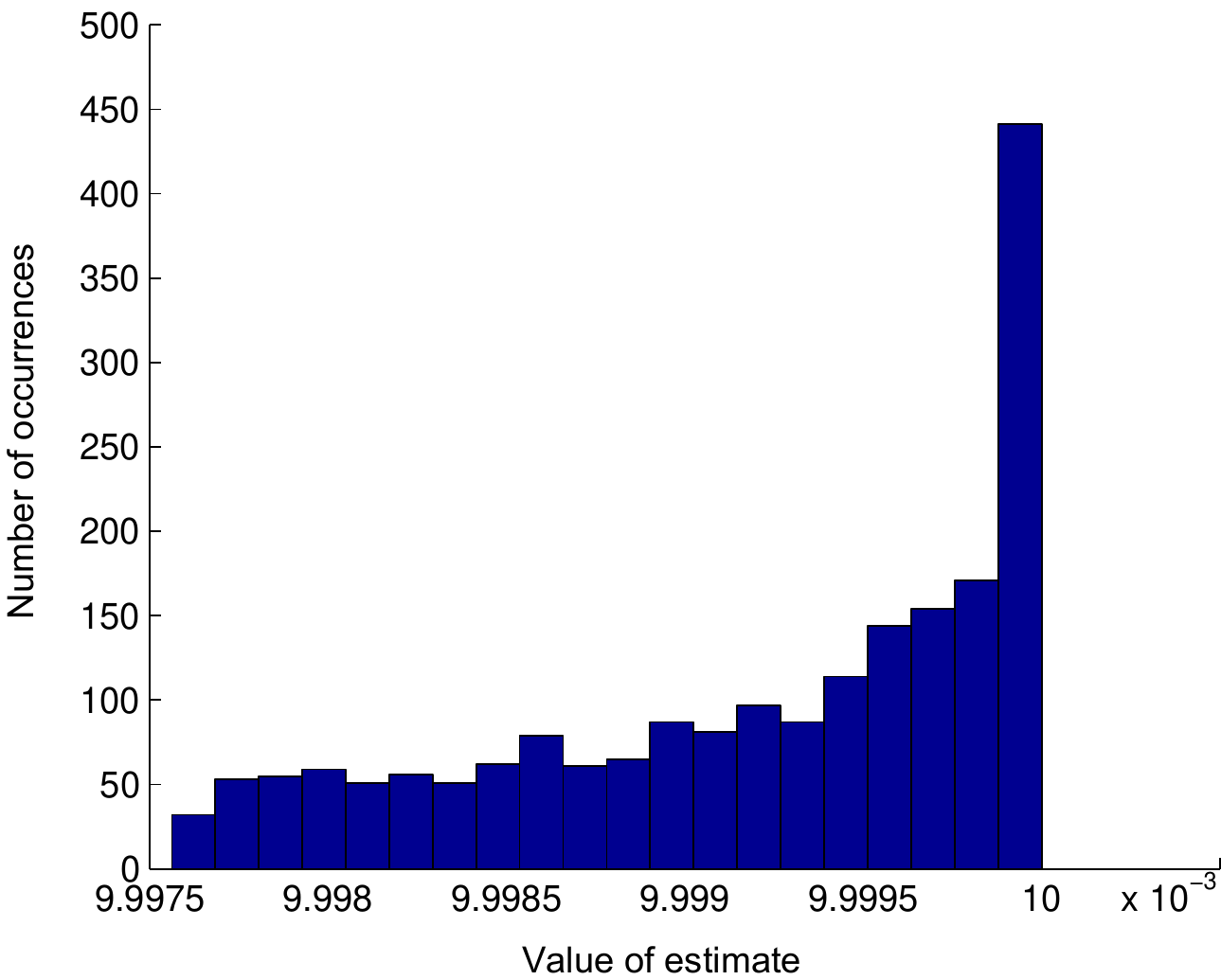}}
\subfigure[]{
\includegraphics[scale=0.6]{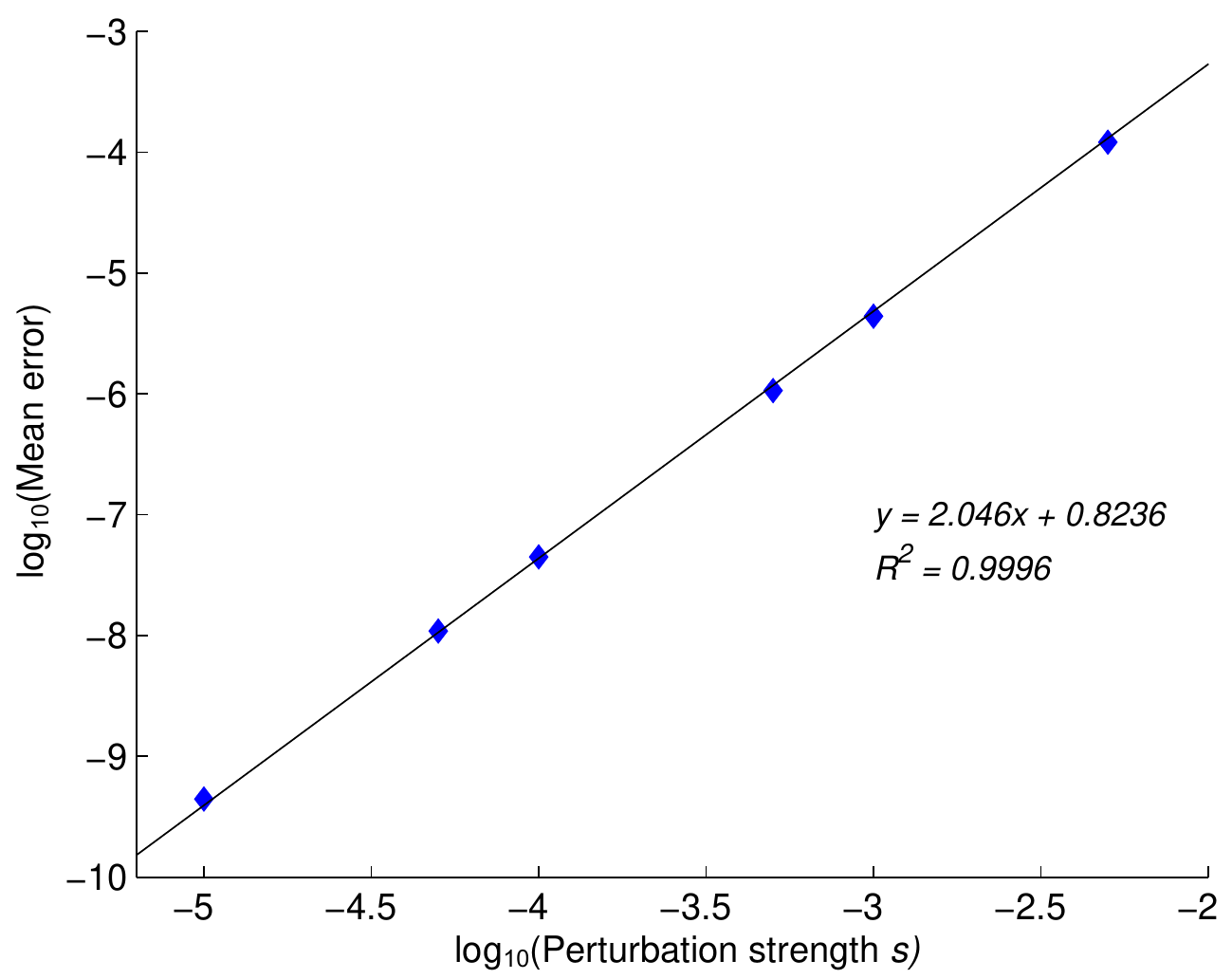}}
\caption{\label{fig:T1numerics} 
a ) Histogram of numerical estimates of $\Gamma_1$ from
2000 independent population inversion experiments using a simulated noisy generalized damping channel with parameters set as follows: $\Gamma_2' = 0.1$, $\Gamma_1 = 0.01$, $\beta, \delta, \alpha_r, \alpha_i \in [-0.001,0.001]$, and $\lambda \in [0.8,1]$ are chosen uniformly at random. SPAM errors $n_1,k\in[0,0.02]$ and $n_2\in[-0.02,0.02]$ are chosen uniformly at random. The estimates are obtained by fitting to 
\cref{eq:T1decay} for 100 measurements (without 
finite measurement statistics) at evenly-spaced times over the interval 
$[0,1/\Gamma_1]$. 
The average estimate is $0.01 - 7.9\times10^{-7}$ and the mean error is $2.3 \times 10^{-5}$. 
b) The mean error in 1,000 simulations of $\Gamma_1$ estimation as a function of the perturbation strength $s$ where noise parameters $\beta, \delta, \alpha_r, \alpha_i \in [-s,s]$, both plotted on logarithmic scales (base 10). Recall that the leading correction term on the $\Gamma_1$ estimate is $\frac{\beta^2}{\Gamma_2^*+\Gamma_1}$. Therefore, for $\Gamma_2'$ set to $0.1$, we expect the order of the error in the estimate to scale as twice the order of the perturbation strength, and increase one order for the division by $ \approx \Gamma_2'$. As displayed on the plot, a linear fit of approximately $y = 2x+1$ fits the data well.   
}
\end{figure*}

In conclusion, the population inversion experiment robustly measures $\Gamma_1$ to second order in the perturbation parameters. The leading correction term on the estimate is dominated by the perturbation of the eigenvalue $\eta_3$:
\begin{align}
\Gamma_1^{\text{est}} \approx \Gamma_1 - \frac{\beta^2}{\Gamma_2' - \Gamma_1}.
\end{align}
We demonstrate this robustness numerically as shown in \cref{fig:T1numerics}. Numerical simulation substantiates both the claim of robustness (Fig. 1a), and the specific leading correction term (Fig. 1b).
Numerical simulations include the state preparation and measurement errors associated with the population inversion experiment.

\section{Dephasing rate ($\Gamma_2'$) Estimation}\label{sec:T2}

\begin{figure*}[h]
\centering
\subfigure[~~Ramsey experiment]{
\includegraphics[scale=0.6]{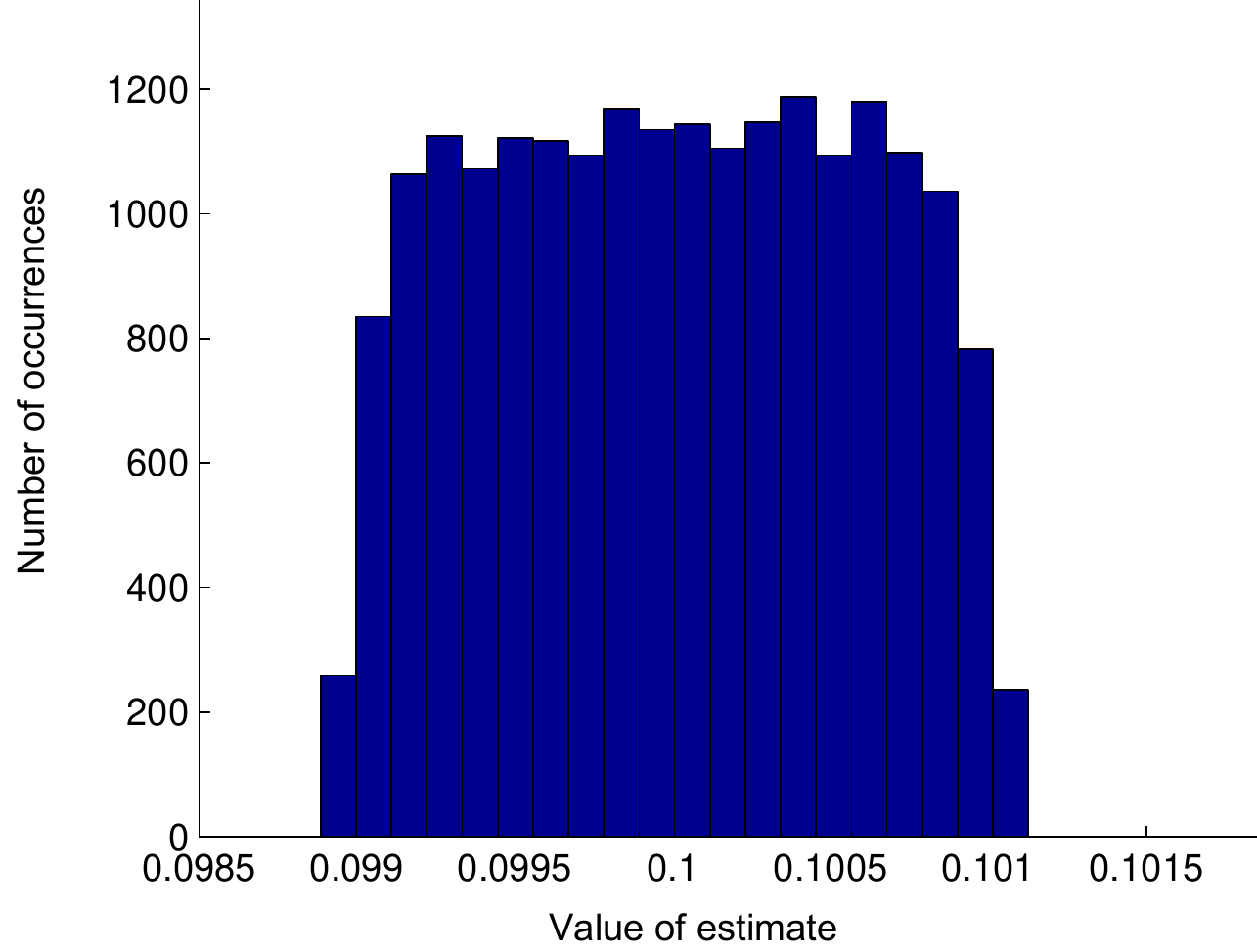}} 
\subfigure[~~Average over $\phi$]{
\includegraphics[scale=0.6]{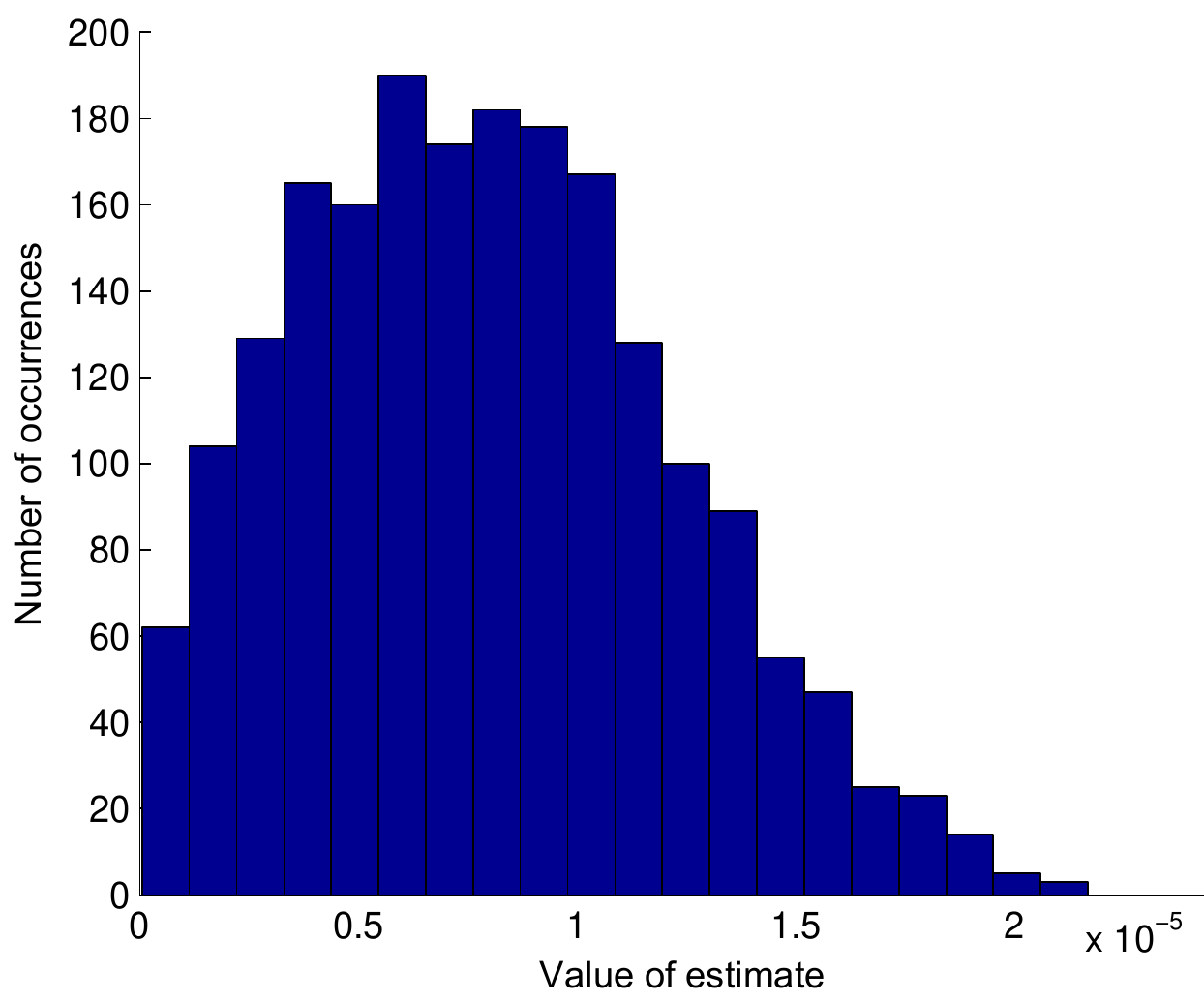}}
\subfigure[~~Comparison of the mean error]{
\includegraphics[scale=0.6]{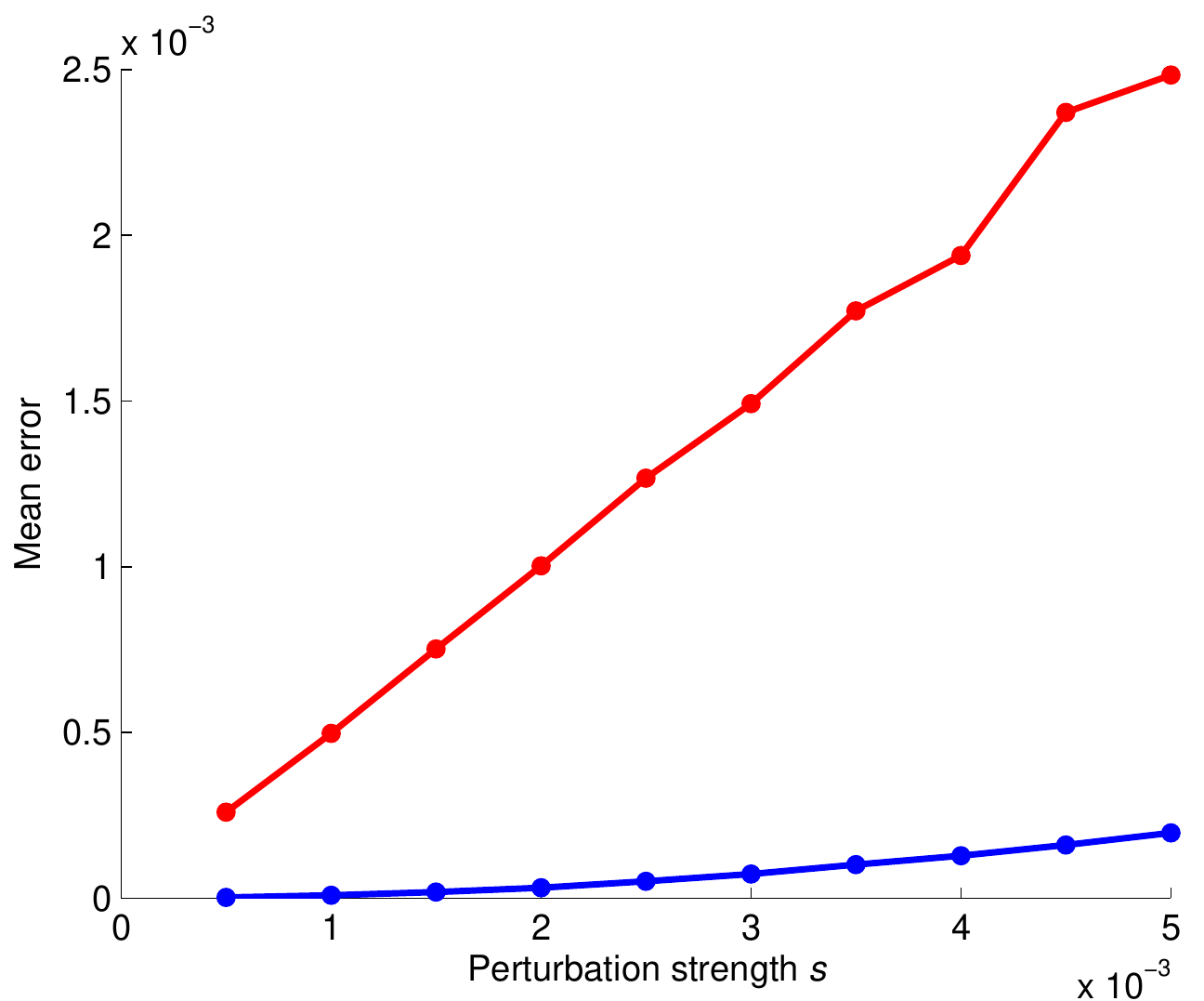}}
\subfigure[~~Order scaling of mean error (average over $\phi$)]{
\includegraphics[scale=0.6]{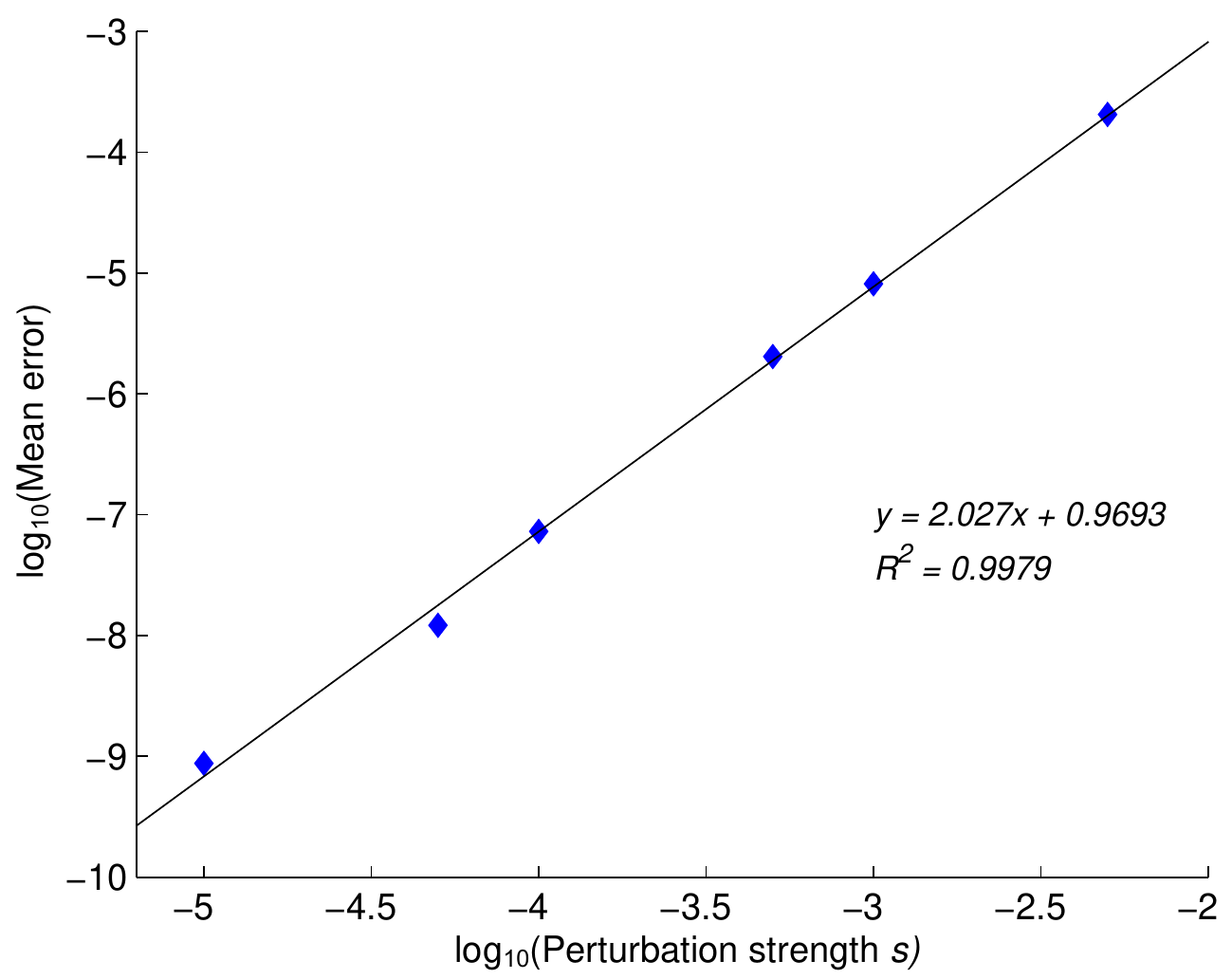}}
\caption{\label{fig:T2numerics} 
a-b) Histograms of numerical estimates of $\Gamma_2'$ from simulations of various fitting experiments. In each simulation, noise parameters are set as follows: $\Gamma_2' = 0.1$, $\Gamma_1 = 0.01$, $\beta, \delta, \alpha_r, \alpha_i \in [-0.001,0.001]$, and $\lambda \in [0.8,1]$ are chosen uniformly at random. SPAM errors $n_1,k\in[0,0.02]$ and $n_2\in[-0.02,0.02]$ are chosen uniformly at random. In all cases, the $\Gamma_2'$ estimates are obtained by fitting the measurement results to \cref{eqn:formQ2sum} for 100 measurements (without finite measurement statistics) at evenly-spaced times over the interval $[0,1/\Gamma_2']$.
a) A histogram of 20,000 numerically obtained estimates for $\Gamma_2'$ where the actual value is set to 0.1, given by a static Ramsey experiment. As expected, the leading contribution to the error in the estimate is given by $|\alpha|$, which has been chosen uniformly at random in the interval $[-.001,.001]$. Tapering of the histogram around the edges of this interval is due to the next leading error terms. The mean error is $5.1\times10^{-4}$. 
b) A histogram of 2,000 numerically obtained estimates of $\Gamma_2'$ where the actual value has been set to 0.1 (histogram shows difference between estimate and actual value), given by an average over preparation and measurement of $xy$ plane states. As expected, the experiment routinely overestimates the true value of $\Gamma_2'$ to second order in the perturbation strength. The mean error is $1.2\times10^{-5}$. 
c) The mean error in 1,000 simulations of $\Gamma_2'$ estimation using a Ramsey experiment (red) compared to the average over plane states (blue) as a function of perturbation strength $s$. For each experiment, $\Gamma_2'$, $\Gamma_1$, $\lambda$, and SPAM parameters are set/selected as above. All perturbation parameters are chosen uniformly at random in the interval $[-s,s]$, defining the perturbation strength. 
d) The mean error in 1,000 simulations of $\Gamma_2'$ estimation using an average over plane states as a function of the perturbation strength, both plotted on logarithmic scales (base 10). Recall that the leading correction term on the $\Gamma_2'$ estimate is $\frac{\beta^2}{\Gamma_2'+\Gamma_1}$. Therefore, for $\Gamma_2'$ set to $0.1$, we expect the order of error in the estimate to scale as twice the order of the perturbation strength, and increase one order for the division by $ \approx \Gamma_2'$. As displayed on the plot, a linear fit of approximately $y = 2x+1$ fits the data well.   
}
\end{figure*}

$\Gamma_2'$ estimation measures the rate of decay of the $xy$ plane of the Bloch sphere. In a static Ramsey experiment, a fixed state in the $xy$ plane is prepared and the resulting decay is measured. Consider this experiment using the $\ket{+}$ state (the $+1$ position of the redefined $x$ axis); after preparation, the state is allowed to precess freely around the $z$ axis for some time $t$, before the ``length" is measured (in the rotating frame) as $Q_2(t) = \Tr(\rho(t)X) = r_x(t)$. This method is not robust to perturbations about the generalized damping channel, as can most easily be seen when $\alpha_r \neq 0$:
\begin{align}
Q_2(t) &= r_x(t)|_{r_x(0)=1} = e^{(\alpha_r - \Gamma_2')t}
\end{align}
Therefore fitting to $Q_2(t)$ does not give an 
accurate estimate of $\Gamma_2'$ in the presence of noise for any 
nonzero value of $\alpha_r$, as is confirmed by the numerical simulations in 
\cref{fig:T2numerics}.

$\Gamma_2'$ estimation may be made robust to perturbations by measuring the decay of a randomly prepared state in the $xy$ plane, and averaging over all such states. This method is effectively already done in practice as the system's frame is generally unknown. Under pure generalized damping, and allowing for SPAM, the expected functional form of such a measurement will be 
\begin{align}
Q_{2,\text{avg}} = c_1e^{-\Gamma_2't} + c_0
\label{eqn:formQ2sum}
\end{align}
which may be fit to extract $\Gamma_2'$.
We now show that this method is robust to perturbations. For $\beta = 0$, the two eigenvectors of $C$ to span the $xy$ plane are 
\begin{align}
v_1 = 
\begin{pmatrix}
\cos\phi \\ \sin\phi \\0 
\end{pmatrix},
v_2 = 
\begin{pmatrix}
-\sin\phi \\ \cos\phi \\0 
\end{pmatrix}
\end{align}
with eigenvalues
\begin{align}
\eta_1 = |\alpha| - \Gamma_2',~~ \eta_2 = -|\alpha| - \Gamma_2' 
\end{align}
where
\begin{align}
\tan\phi = \frac{\alpha_i}{\alpha_r + \sqrt{\alpha_i^2 + \alpha_r^2}}.
\end{align}
Let a general preparation state and its measurement be defined as
\begin{align}
\bs{r}(0) &= \begin{pmatrix} \cos\omega \\ \sin\omega \\ 0\end{pmatrix} & M = \cos\omega X + \sin\omega Y 
\label{eqn:phi_experiment}
\end{align}
then
\begin{align}
Q_{2}(\omega) & = \cos\omega r_x(t) + \sin\omega r_y(t) \notag \\
& = C_1\lambda_1/\eta_1(e^{\eta_1t}-1) + C_1^2e^{\eta_1t} \notag \\
&+ C_2\lambda_2/\eta_2(e^{\eta_2t}-1) + C_2^2e^{\eta_2t}
\end{align}
where
\begin{equation}
\begin{aligned}[c]
\lambda_1& = 2\sqrt{2}\delta\cos\phi \\
\lambda_2& = -2\sqrt{2}\delta\sin\phi 
\end{aligned}
\enskip
\begin{aligned}[c]
C_1 &= \cos\phi\cos\omega +\sin\phi\sin\omega \\
C_2 &= -\sin\phi\cos\omega +\cos\phi\sin\omega. 
\end{aligned}
\end{equation}
Averaging over all $\omega$,
\begin{align}
\label{eqn:Q2avg_noB}
Q_{2,\text{avg}} = \frac{1}{2\pi} \int_{0}^{2\pi}Q_{2}(\omega) d\omega = \frac{1}{2}(e^{\eta_1t} + e^{\eta_2t}).
\end{align}
Fitting the measured quantity \cref{eqn:Q2avg_noB} to functional form \cref{eqn:formQ2sum} will measure the average of the eigenvalues $\eta_1$ and $\eta_2$, which in this case is $-\Gamma_2'$. Therefore an experimental average over $xy$ plane states will estimate $\Gamma_2'$ robustly to perturbations in $\alpha$ and $\delta$.  

For $\beta \neq 0$, the $\Gamma_2'$ estimate will pick up the $\Gamma_1$ decay on the order of $\beta^2$, due to the non-zero overlap of the $xy$ plane with the third eigenvector $v_3$. For example, letting $\alpha_i = 0$ for simplicity, we obtain for the averaged measurement
\begin{align}
Q_{2,\text{avg}} = \frac{1}{2}(\cos^2\theta e^{\eta_1t} + e^{\eta_2t} + \sin^2\theta^{\eta_3t}) 
\end{align}  
and therefore the estimated value will be the average of $\eta_1$ and $\eta_2$ with additional contributions from $\eta_1$ and $\eta_3$ on the order of $\beta^2$ (recall $\cos(\theta) \approx 1 + O(\beta^2)$, $\sin(\theta) \in O(\beta)$). 
Non-zero $\alpha_i$ primarily has the effect of rotating $v_1$ and $v_2$ around the $z$-axis of the Bloch sphere, which is being averaged over. 
The average of the eigenvalues $\eta_1$ and $\eta_2$ is approximately given by $-\Gamma_2' - \frac{1}{2}\beta^2/(\Gamma_2' - \Gamma_1)$ (see Appendix D).
Therefore, the eigenvector contributions are negligible compared to the first leading correction term due to the eigenvalue perturbations. 

In conclusion, the Ramsey experiment averaged over plane states robustly measures $\Gamma_2'$ to second order in the perturbation parameters. The leading correction term on the estimate is given by 
\begin{align}
\Gamma_2'^{\text{est}} \approx \Gamma_2' + \frac{1}{2}\frac{\beta^2}{\Gamma_2' - \Gamma_1}.
\end{align}
Numeric simulation shows that the estimate behaves as derived here (see \cref{fig:T2numerics}).

\section{Predicted outcomes of randomized benchmarking}

In this section we provide the error rate and unitarity \cite{Wallman:2015Aug} expected in a randomized benchmarking experiment attributed to the ideal generalized damping channel, which serve as predicted values in the case when error consists of only of the generalized damping processes. 


The error rate obtained from a randomized benchmarking experiment may be calculated as $r = 1-\mathcal{F}$, where $\mathcal{F}$ is the average fidelity
$ \mathcal{F} = \frac{1}{6}\sum_i\bra{\psi_i}\mathcal{E}(\psi_i)\ket{\psi_i} $ 
and $\ket{\psi_i}$ are the eigenstates of $X,Y,Z$ operators. For the 
generalized damping channel,
\begin{align}
\begin{split}
r_{\text{ideal}}& = \frac{1}{2} - \frac{1}{6}e^{-\Gamma_1\Delta t}-\frac{1}{3}e^{-(\Gamma_1/2+\Gamma_2)\Delta t}
\\& = \frac{1}{3}+\frac{\gamma_1}{6} -\frac{1}{3}\sqrt{(1-\gamma_1)(1-\gamma_2)}.
\end{split}
\end{align}
The second line follows from the discrete-time representation of the generalized damping channel (see appendix A), where $\gamma_1 = 1-e^{-\Gamma_1\Delta t}$, $\gamma_2 = 1-e^{-2\Gamma_2\Delta t}$, and $\Delta t$ is the average gate duration.

Additionally, we can obtain more benchmark values if we perform the randomized benchmarking procedure with sequences of random Pauli gates and an additional projection step, as in \cite{Dugas:2015}. The error rate $r^\sigma$ represents the value obtained with the projection $\frac{1}{2}(I + \sigma)$ applied at the end of the sequence where $\sigma \in \{X,Y,Z\}$: 
\begin{align}
r^X_{\text{ideal}} = r^Y_{\text{ideal}} &= \frac{1}{2} - \frac{1}{6}\sqrt{(1-\gamma_1)(1-\gamma_2)} \notag \\ 
r^Z_{\text{ideal}} &= \frac{1}{2} - \frac{1}{6}(1-\gamma_1).
\end{align}
Similarly, the ideal unitarity is given by 
\begin{align}
\mathcal{U}_{\text{ideal}} = \frac{1}{3}\text{Tr}(\Lambda_u^\dagger \Lambda_u) = \frac{1}{3}(3-4\gamma_1 -2\gamma_2 +2\gamma_1\gamma_2+\gamma_1^2)
\end{align} 
where $\Lambda_u$ is the unital part of the generalized damping channel. Together these values provide a benchmark to check whether noise is limited by damping processes. In reality, these ideal values cannot be observed because of the additional control errors introduced by the randomized benchmarking procedure. In the following section, we will use the differences between the ideal and actual measured values to bound the diamond distance of the noise channel from the identity. 

\section{Comparing experiments to a threshold}

The diamond distance of a noise channel from the identity channel is the relevant quantity for comparison to fault-tolerant thresholds \cite{Kitaev:1997}, defined as 
\begin{align}
D(\Lambda) = \frac{1}{2}||I - \Lambda||_\diamond = \frac{1}{2}\text{sup}_\rho||I \otimes (I - \Lambda)(\rho)||_1
\end{align}
The diamond distance cannot be estimated by any direct experimental method unless the noise is known to be a Pauli channel (achievable via randomized compiling \cite{Wallman:2016}), in which case the diamond norm may be estimated through randomized benchmarking \cite{Magesan:2012}. However in the absence of an error suppression scheme, such as randomized compiling, experimental estimation techniques for the diamond distance are unknown. Therefore it is common practice to provide an upper bound on the diamond distance in terms of known experimental quantities.  

As a starting point, we obtain an upper bound (see appendix B) for the diamond distance between the identity and the perfect generalized damping channel $D_{GD} = \frac{1}{2}||I - \Lambda_{GD} ||_\diamond$ as 
\begin{align}
D_{GD} \leq \epsilon^{\textrm UB}_{GD} = \frac{1}{2} \Big[1 - \sqrt{(1-\gamma_1)(1-\gamma_2)} - \frac{1}{2}\gamma_1 + 2\lambda\gamma_1 \Big].
\label{eqn:D1}
\end{align}
This bound is valid when noise is perfectly described by a generalized damping channel, which is an idealization that will not hold true in many realistic experiments. Note that the bound does not scale as $\sqrt{r}$, as desired.  

Next we obtain a usable bound that accounts for perturbative deviations from the ideal generalized damping channel. The bound requires only parameters that can be estimated via the robust experimental procedures analyzed above; in particular the usable bound is based on the differences between the actual measurements of the error rates $r^\sigma$ and unitarity $\mathcal{U}$ and their ideal values predicted from the estimated damping parameters. We assume 
the total error channel may be written (in super-operator representation) as
\begin{align}\label{eqn:pertChannel}
\Lambda = \Lambda_{GD} + \begin{pmatrix} 0 & 0 \\ 0 & \mathcal{E} \\ \end{pmatrix}
\end{align}
where we have assumed the total channel to be trace-preserving and contain no leakage errors. We further assume that the perturbation contains only unital components, where $\mathcal{E}$ represents the unital block. 

Note that \cref{eqn:pertChannel} is a slightly different perturbation model than was described by \cref{eq:general_A}. In the previous perturbation analysis, we were interested in measuring the intrinsic damping parameters in the presence of additional, unknown, intrinsic errors (captured by the dissipator terms). The robust experimental procedures for this estimation did not introduce additional control errors beyond state preparation and measurement errors. However we now wish to compare the ideal and measured values from randomized benchmarking experiments to build our upper bound on the diamond distance.    
Implementing gates required for the randomized-benchmarking type experiments can add two types of noise: imperfect Hamiltonians (systematic calibration errors) and dissipation terms (oscillating control noise). 
Therefore the current perturbation model must allow for both Hamiltonian terms and dissipator terms, whereas the previous continuous-time perturbation model allowed only dissipator terms. 

We can bound the diamond norm of $\Lambda$ from the identity as 
\begin{align}
||I - \Lambda||_\diamond \leq \epsilon^{\textrm UB} = 2\epsilon^{\textrm UB}_{GD} + \sum_{i,j}|\mathcal{E}_{i,j}|
\end{align}
which is obtained by splitting $\Lambda$ into $\Lambda_{GD}$ plus a sum of one element matrices and applying the triangle inequality. The diamond norm of a matrix with a single non-zero entry in the unital block is given by the absolute value of that element as shown in appendix C.
Applying the Cauchy-Schwarz inequality 
\begin{align}\label{eqn:bound1}
||I - \Lambda||_\diamond &\leq 2\epsilon^{\textrm UB}_{GD} + \sqrt{9\sum_{i,j}\mathcal{E}_{i,j}^2} \\ \notag
&= 1 - \sqrt{(1-\gamma_1)(1-\gamma_2)} + \frac{3}{2}\gamma_1 + \sqrt{9\sum_{i,j}\mathcal{E}_{i,j}^2}
\end{align}
where we have replaced the unknown $\lambda$ in $\epsilon^{\textrm UB}_{GD}$ by its upper bound of $1$ since it may not be robustly estimated.
Finally we estimate $\sum_{i,j}\mathcal{E}_{i,j}^2$ from comparing the measured values of $r^\sigma$ and $\mathcal{U}$  to the ideal values:
\begin{align}\label{eqn:Esum}
\begin{split}
\sum_{i,j}\mathcal{E}_{i,j}^2 = &~3(\mathcal{U} - \mathcal{U}_{\text{ideal}}) -12(1-\gamma_1)(r_{\text{ideal}}^Z- r^Z)
\\& -12\sqrt{(1-\gamma_1)(1-\gamma_2)}(r_{\text{ideal}}^X- r^X + r_{\text{ideal}}^Y- r^Y). 
\end{split}
\end{align}
Therefore, comparing unitarity and fidelity to their expected values quantifies both the dissipation terms and imperfect Hamiltonians.

Calculation of the upper bound defined by \cref{eqn:bound1} requires knowledge of the values of $\gamma_1$ and $\gamma_2$. As discussed, these parameters may be robustly estimated to second order in the perturbation terms, where the leading correction term on each is $O(\beta^2)$. Therefore, the user calculated value for bound \cref{eqn:bound1} differs from its actual value by $O(\beta^2)$. Appendix F provides a full expression for the error in the user calculated bound, denoted simply by $O(\beta^2)$ here for brevity. If perturbations from the generalized damping channel are small, this discrepancy is negligible.

In the case that the $\beta^2$ cannot be assumed to be negligible compared to the generalized damping parameters, we provide a robust bound which compensates for the unknown error terms:
\begin{align} \label{eqn:bound_robust}
\epsilon^{\textrm UB}_{\text{robust}} &= 2\epsilon^{\textrm UB}_{GD} + 12(r_\text{ideal}^X + r_\text{ideal}^Y - r^X - r^Y) \notag \\
&+ \sqrt{9\sum \mathcal{E}^2_{\text{robust}}} \notag \\
\sum \mathcal{E}^2_{\text{robust}} &= \sum_{i,j} \mathcal{E}_{i,j}^2 +6(r_\text{ideal}^X + r_\text{ideal}^Y - r^X - r^Y).
\end{align}
The robust bound is constructed in such a way that the user calculated robust bound will be larger than the exact bound \cref{eqn:bound1}. More details on the construction of the robust bound can be found in Appendix F. In this way, we eliminate the assumption of negligible second order noise terms, and provide a true robust bound. 

\begin{figure*}[h]
\centering
\subfigure[~~Scaling of bounds by the order of the perturbation strength]{
\includegraphics[scale=0.6]{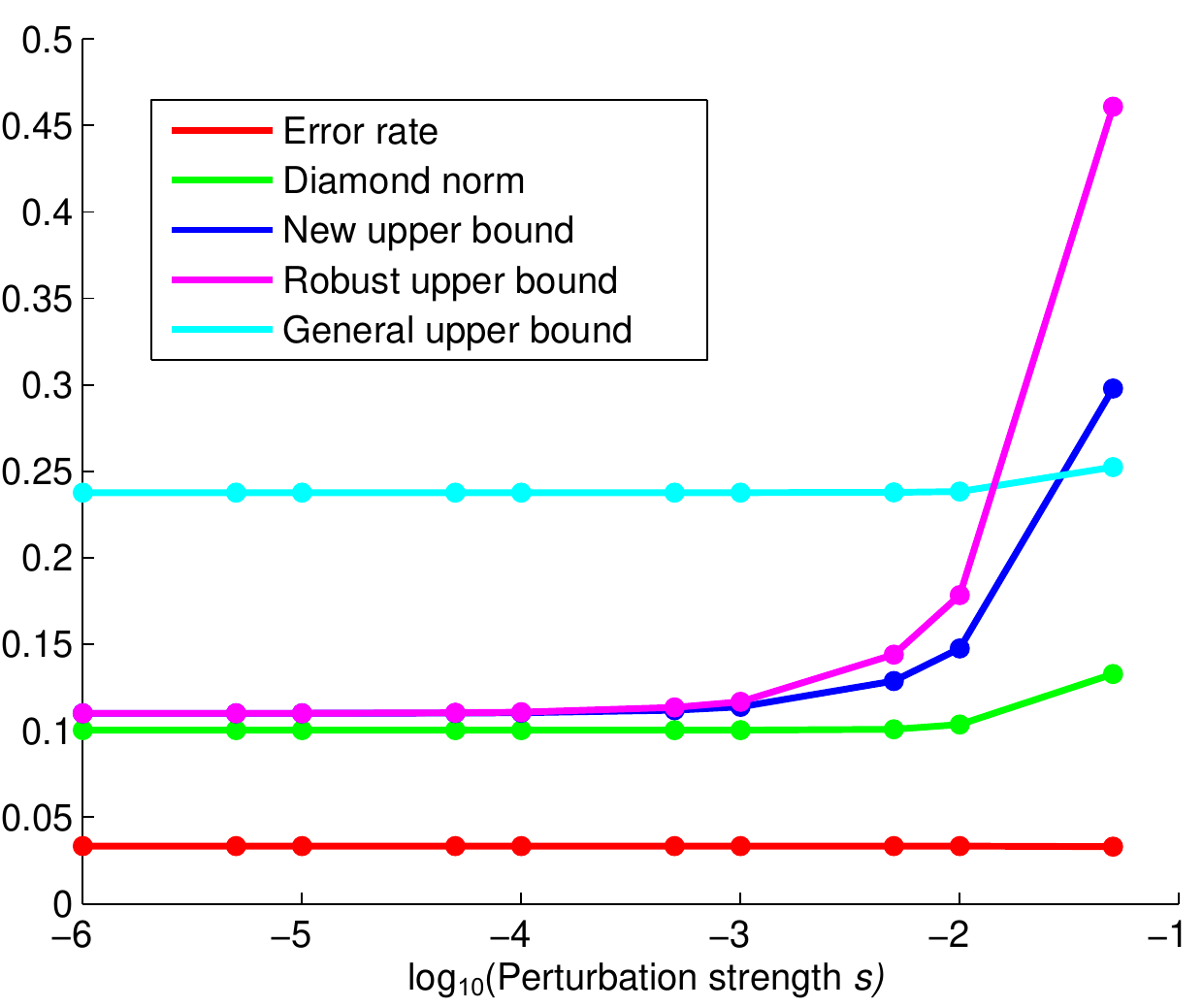}} 
\subfigure[~~Region of usefulness of the robust bound by the order of the error rate]{
\includegraphics[scale=0.6]{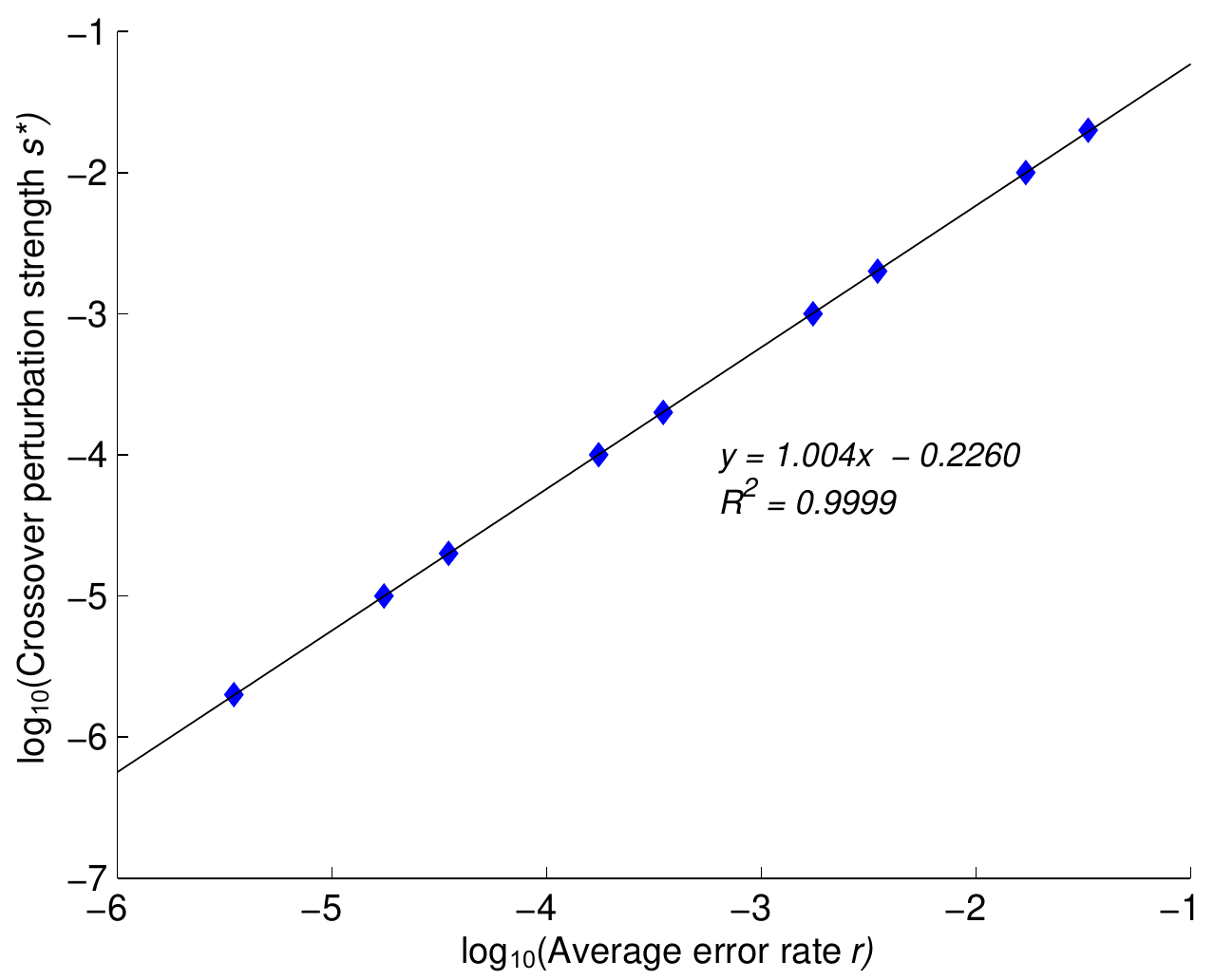}} 
\caption{\label{fig:DNnumerics_robust} 
a) The average error rate, diamond norm from the identity channel, and upper bounds on the diamond norm (general given by \cite{Wallman:2015Nov}, new given by \cref{eqn:bound1}, and robust given by \cref{eqn:bound_robust}) averaged over 5000 simulated noise channels where the generalized damping parameters have been set as $\Gamma_2' = 0.1$, $\Gamma_1 = 0.01$, and $\lambda \in [0.8,1]$. The average gate duration has been set to $t = 1 = 10\Gamma_2'$ ($\gamma_1 \approx \Gamma_1$, $\gamma_2 \approx \Gamma_2$). Perturbation parameters $\alpha, \beta, \delta$ have been selected uniformly at random in the interval $[-s,s]$, which defines the perturbation strength $s$ as the absolute size of the perturbation parameters. Note that here we do use the perturbation model \cref{eqn:C}, on which the error in the estimates and subsequent construction of the robust bound are based. The new upper bound \cref{eqn:bound1} and the robust upper bound \cref{eqn:bound_robust} are comparable for small perturbation strengths and provide a significant improvement over the general bound. The robust upper bound provides an improvement over the general bound while the perturbation parameters are less than $s^* = 10^{-2}$, which is $O(r)$ in this case ($r \approx .033$ is essentially constant due to the fixed generalized damping parameters).   
b) The crossover perturbation strength $s^*$ above which the robust bound no longer provides an improvement over the general bound as a function of the average error rate $r$. Generalized damping parameters are set as $\Gamma_1 = \Gamma_2'/10$, $\lambda \in [.8,1]$, and $\Gamma_2'$ is varied to achieve the varying error rate. Each point is averaged over 10,000 random noise channels. The linear fit and slope of 1 indicates that the crossover perturbation strength is consistently first order in the error rate.  
}
\end{figure*}

Also note that $||I - \Lambda||_\diamond$ can be bounded in terms of the measured $r$ and $\mathcal{U}$ directly as in \cite{Wallman:2015Nov}. Assuming that the perturbation from the generalized damping channel is small, both the bound \cref{eqn:bound1} and the robust bound \cref{eqn:bound_robust} provide a significant improvement over this general bound, as shown in the numerical simulations \cref{fig:DNnumerics_robust}. (Also see \cref{fig:DNnumerics} appendix H). 

\section{Summary}
In summary, we have provided an in-depth analysis of how commonly measured 
experimental quantities should behave for a channel close to generalized 
damping. 
We have shown that the damping rate parameters $\Gamma_1$ and $\Gamma_2'$ may be estimated robustly to second order in the perturbation parameters, and provide the leading error in the estimates, which is shown to be $O(\beta^2)$. 
We provide expressions for the ideal values of numerous randomized benchmarking experiments, which may be compared to the actual measured values in order to better understand if noise is limited by damping processes. 
The differences between actual and ideal values may be used to bound the diamond distance, and this bound provides a way to make meaningful statements about fault-tolerant threshold theorems. We provide two variants of this bound: (i) one that can be used with the assumption that $O(\beta^2)$ terms are negligible compared to the generalized damping parameters, and (ii) a more robust version that does not require this assumption. 
Both bounds improve over the general bound for channels close to generalized damping, however bound (i) provides a more significant improvement. Therefore future work may be to develop methods of experimentally verifying this assumption. For example, the development of randomized-benchmarking-type experiments that can robustly estimate $\beta$.    
Additionally, we emphasize that the bound \cref{eqn:bound1} is true for any model of unital perturbation, and \cref{eqn:Esum} is true for any perturbation model, where both are made usable by the robust experimental estimation of $\gamma_1$ and $\gamma_2$. Therefore future work may be to argue this robustness (and find leading error terms) for more specific noise models relevant to various experimental systems, which will also likely improve the subsequent construction of a robust bound. 

\textit{Acknowledgments $-$} 
We thank Steve Flammia and David Cory for helpful conversations.
This research was undertaken thanks in part to funding from the Canada First Research Excellence
Fund, the Government of Ontario, the Government of Canada through NSERC, and Army Research Office grant W911NF-14-1-0103.

\clearpage
\onecolumngrid
\appendix

\section{Discrete-time representations of the generalized damping channel}\label{app:A}

An equivalent discrete-time channel giving the same evolution as 
\cref{eqn:GD_ME} is
\begin{align}
\mathcal{E}_\mathrm{GD}(\rho) = \sum_j K_j\rho K_j^\dagger
\end{align}
with Kraus operators
\begin{align}
K_0 &=\sqrt{\frac{1-\gamma_1/2-x}{a^2+b^2}}(a\op{0}-b\op{1})
\notag\\
K_1 &= \sqrt{\lambda\gamma_1}\op{0}{1} \notag\\
K_2 &=\sqrt{\frac{1-\gamma_1/2+x}{a^2+b^2}}(b\op{0}+a\op{1})
\notag\\
K_3 &= \sqrt{(1-\lambda)\gamma_1}\op{1}{0} 
\end{align}
where $\gamma_1 = 1-e^{-\Gamma_1 \Delta t}$, $\gamma_2 = 1-e^{-2\Gamma_2 \Delta t}$, $x = [(1-\gamma_1)(1-\gamma_2)-\gamma_1^2\lambda(1-\lambda)+1/4\gamma_1^2]^{1/2}$, $a = x+\gamma_1/2-\gamma_1\lambda$, and $b = \sqrt{(1-\gamma_1)(1-\gamma_2)}$. 

In Pauli-Liouville representation, the generalized damping channel is given by
\begin{align}
\Lambda_{GD} =
\begin{pmatrix}
1& 0& 0& 0 \\
0 & \sqrt{(1-\gamma_1)(1-\gamma_2)} & 0 & 0 \\
0 & 0 & \sqrt{(1-\gamma_1)(1-\gamma_2)} & 0 \\
\gamma_1(2\lambda-1) & 0 & 0 & 1-\gamma_1 \\
\end{pmatrix}.
\end{align}

The perturbed generalized damping model \cref{eqn:C} can be described by the discrete time process matrix over a time interval $\Delta t$
\begin{align}
\Lambda_{PGD} =
\begin{pmatrix}
1& 0& 0& 0 \\
n_x & p_x & 0 & 0 \\
n_y & 0 & p_y & 0 \\
n_z & 0 & 0 & p_z \\
\end{pmatrix}
\end{align}
where
\begin{align}
n_j = \lambda_j(e^{\eta_j \Delta t} - 1)/\eta_j \\ \notag
p_j = e^{\eta_j \Delta t}
\end{align}
and we are working in the orthonormal basis $\{I, V_1, V_2, V_3\}$ and $V_j = v_j \cdot \{X,Y,Z\}$ which follows from \cref{eq:evolved_bloch} exactly. Note that perturbations to the continuous time generalized damping model, as described by \cref{eqn:C} result in stochastic noise.

\section{Diamond distance bounds for the generalized damping channel}\label{app:B}
The diamond distance $D_{GD} = \frac{1}{2}||I - \Lambda_{GD} 
||_\diamond$ is defined as 
\begin{align}
D_{GD} = \frac{1}{2} \text{sup}_{\rho} || I \otimes (I-\Lambda_{GD})(\rho)||_1.
\end{align}
We may split $I-\Lambda_{GD}$ into its diagonal and non-diagonal part $I-\Lambda_{GD} = \mathcal{E}_1 + \mathcal{E}_2$
\begin{align}
\mathcal{E}_1 = \begin{pmatrix} 0 & 0 & 0 & 0 \\ 0 & 
1-\sqrt{(1-\gamma_1)(1-\gamma_2)} & 0 & 0 \\ 0 & 0 & 
1-\sqrt{(1-\gamma_1)(1-\gamma_2)} & 0 \\ 0 & 0 & 0 & \gamma_1 \\ 
\end{pmatrix}~~ \mathcal{E}_2 = \begin{pmatrix} 0 & 0 & 0 & 0 \\ 0 & 0 & 0 & 0 
\\ 0 & 0 & 0 & 0 \\ -\gamma_1(2\lambda-1) & 0 & 0 & 0 \\ \end{pmatrix} 
\end{align}
and then apply the triangle inequality $||I - \Lambda_{GD} ||_\diamond \leq 
||\mathcal{E}_1||_\diamond + ||\mathcal{E}_2||_\diamond$. 
Since $\mathcal{E}_1$ is a Pauli channel, we can express its action as $\mathcal{E}_1(\rho) = \sum_i v_iP_i\rho P_i^\dagger$ where $P_i$ are the single qubit Pauli operators $\{I,X,Y,Z\}$ and 
\begin{align}
\vec{v} = \begin{pmatrix}1/2(1-\sqrt{(1-\gamma_1)(1-\gamma_2)})+\gamma_1/4 \\
-\gamma_1/4 \\
-\gamma_1/4 \\
\gamma_1/4 - 1/2(1-\sqrt{(1-\gamma_1)(1-\gamma_2)}) \end{pmatrix}.
\end{align}
The diamond norm of a Pauli channel is directly computable as $\sum_i |v_i|$ \cite{Magesan:2012}, therefore
\begin{align}
||\mathcal{E}_1||_\diamond = 1 - \sqrt{(1-\gamma_1)(1-\gamma_2)} + \frac{1}{2}\gamma_1.
\end{align}
Next $||\mathcal{E}_2||_\diamond = \sup_{\rho}||I \otimes 
\mathcal{E}_2(\rho)||_1$ where $\rho$ can be assumed to be a pure state, $\rho 
= \op{\psi}$ and $\ket{\psi} = (a~b~c~d)^T$. Working this out, we 
get: 
\begin{align}
I \otimes \mathcal{E}_2(\rho) = \frac{-\gamma_1(2\lambda-1)}{2}\begin{pmatrix} 
|a|^2 + |b|^2 & 0 & ac^*+bd^* & 0 \\ 0 & -(|a|^2 + |b|^2) & 0 & -(ac^*+bd^*) \\ 
ca^*+db^* & 0 & |c|^2 + |d|^2 & 0 \\ 0 & -(ca^*+db^*) & 0 & -(|c|^2 + |d|^2) \\ 
\end{pmatrix} 
\end{align}
which has eigenvalues $\lambda = \pm \frac{\gamma_1(2\lambda-1)}{4} [1 \pm \sqrt{1-4(ad-bc)(a^*d^* - b^*c^*)}]$ so that for any $a,b,c,d$, $\sum |\lambda_k| = \gamma_1(2\lambda-1)$ and therefore $||\mathcal{E}_2||_\diamond = \gamma_1(2\lambda-1)$. Putting things together, the upper bound is 
\begin{align}
||I - \Lambda_{GD} ||_\diamond \leq 1 - \sqrt{(1-\gamma_1)(1-\gamma_2)} - \frac{1}{2}\gamma_1 + 2\lambda\gamma_1 .
\end{align}

\section{The diamond norm of single element matrices}\label{app:C}
Here we show that the diamond norm of a matrix $M = \begin{pmatrix} 0&0\\ \alpha&\mathcal{E}\\ \end{pmatrix}$ with a single non-zero element $x$ in either the unital block $\mathcal{E}$ or non-unital vector $\alpha$ is equal to $|x|$. 
Due to the invariance of the diamond norm under unitary conjugation, we need to show this for only two cases: when $x$ is some element of $\alpha$, and when $x$ is some element of $\mathcal{E}$. 

\subsection{$x$ is an element of $\alpha$}
Let $\alpha = \begin{pmatrix} 0\\0\\x\\ \end{pmatrix}$, $\mathcal{E} = 0$, and 
$\rho = \op{\psi}$ where $\ket{\psi} = (a~b~c~d)^T$. Then 
$||M||_\diamond = \sup_{\rho}||I \otimes M(\rho)||_1$. 
\begin{align}
I \otimes M(\rho) = \frac{x}{2}\begin{pmatrix} |a|^2 + |b|^2 & 0 & ac^*+bd^* & 
0 \\ 0 & -(|a|^2 + |b|^2) & 0 & -(ac^*+bd^*) \\ ca^*+db^* & 0 & |c|^2 + |d|^2 & 
0 \\ 0 & -(ca^*+db^*) & 0 & -(|c|^2 + |d|^2) \\ \end{pmatrix} 
\end{align}
which has eigenvalues $\lambda = \pm \frac{x}{4} [1 \pm \sqrt{1-4(ad-bc)(a^*d^* - b^*c^*)}]$ so that for any $a,b,c,d$, $\sum |\lambda_k| = |x|$ and therefore $||M||_\diamond = |x|$.

\subsection{$x$ is an element of $\mathcal{E}$}
Let $\mathcal{E} = \begin{pmatrix} 0&0&0\\0&0&0\\0&0&x\\ \end{pmatrix}$ and 
$\alpha = 0$.  Note that $M$ is a Pauli channel, $M(\rho) = \sum_i v_i P_i \rho P_i^\dagger$, where $\vec{v} = 1/4(x,-x,-x,x)$. Therefore $||M||_\diamond = \sum_i |v_i| = |x|$.

\section{Exact eigenvalues of $C$}
The governing matrix of evolution of the Bloch sphere for the perturbed generalized damping model is given by 
\begin{align}
C &= \left(
\begin{array}{ccc}
 \alpha_r-\Gamma_2' & \alpha_i & \beta  \\
 \alpha_i & -\alpha_r-\Gamma_2' & 0 \\
 \beta  & 0 & -\Gamma_1 \\
\end{array}
\right)
\end{align}
where all elements are real. The eigenvalues of a $3\times3$ real symmetric matrix are given by \cite{Smith:1961}
\begin{align}
\eta_1 &= m - \sqrt{p}(\cos\phi -\sqrt{3}\sin\phi)   \notag \\
\eta_2 &= m - \sqrt{p}(\cos\phi +\sqrt{3}\sin\phi)  \notag \\
\eta_3 &= m +2\sqrt{p}\cos\phi
\end{align}
where 
\begin{align}
\phi &= \frac{1}{3}\tan^{-1} \frac{\sqrt{p^3-q^2}}{q} \\ \notag
m &= \frac{1}{3}\text{tr}(C) = \frac{-1}{3}(\Gamma_1+2\Gamma_2') \\ \notag
q &= \frac{1}{2}\text{det}(C-mI) = -(\Gamma_1/3 - \Gamma_2'/3)^3 + (\alpha_r^2 + \alpha_i^2 - 1/2\beta^2)(\Gamma_1/3-\Gamma_2'/3) + \beta^2\alpha_r/2 \\ \notag
p &= \frac{1}{6}\sum(C-mI)_i^2 = (\Gamma_2'/3 - \Gamma_1/3)^2 + \alpha_r^2/3 + \alpha_i^2/3 +\beta^2/3.
\end{align}
Assuming that the perturbation terms are small compared to the generalized damping parameters, and using $\Gamma_2' \gg \Gamma_1$, we find the approximations
\begin{align}
\eta_1 &\approx -\Gamma_2' + |\alpha| - \frac{\beta^2}{\Gamma_2' - \Gamma_1} \notag \\
\eta_2 &\approx -\Gamma_2' - |\alpha| \notag \\
\eta_3 &\approx -\Gamma_1 + \frac{\beta^2}{\Gamma_2' - \Gamma_1}.
\end{align}

\section{A note on the error rate of the perturbed generalized damping model}
The error rate of the generalized damping channel is given by
\begin{align}
r_{GD} = \frac{1}{2} - \frac{1}{3}e^{-t\Gamma_2'} - \frac{1}{6}e^{-t\Gamma_1}.
\end{align}
Non-zero perturbation terms as defined by \cref{eqn:C} have the effect of decreasing the error rate. For demonstration purposes, we can look at a couple simple examples. Let $\alpha_i, \alpha_r = 0$ and $\beta,\delta \neq 0$, then 
\begin{align}
r = \frac{1}{2} - \frac{1}{6}e^{-\frac{t}{2}(\Gamma_2' + \Gamma_1 + \sqrt{4\beta^2 + (\Gamma_2'-\Gamma_1)^2}}- \frac{1}{6}e^{-t\Gamma_2'} - \frac{1}{6}e^{-\frac{t}{2}(\Gamma_2' + \Gamma_1 - \sqrt{4\beta^2 + (\Gamma_2'-\Gamma_1)^2}}  < r_{GD}.
\end{align}
Alternatively, let $\alpha_i, \alpha_r \neq 0$ and $\beta,\delta = 0$, then
\begin{align}
r = \frac{1}{2} - \frac{1}{6}e^{-t(\Gamma_2' - |\alpha|)}- \frac{1}{6}e^{-t(\Gamma_2' + |\alpha|)} - \frac{1}{6}e^{-t\Gamma1} < r_{GD}.
\end{align}
In full generality, 
\begin{align}
r = \frac{1}{2} - \frac{1}{6}(e^{\eta_1t} + e^{\eta_2t} + e^{\eta_3t}) \leq r_{GD}
\end{align}
where $\eta_i$ are the eigenvalues of $C$ (exact expressions given in Appendix D).

\section{The user calculated bound on the diamond norm}
The upper bound on the diamond norm of the perturbed generalized damping channel from the identity is given by \cref{eqn:bound1} where $\gamma_1$ and $\gamma_2$ represent their true values. As discussed, $\gamma_1$ and $\gamma_2$ may be robustly estimated to second order in the unknown perturbation parameters. Therefore the user calculated version of the bound \cref{eqn:bound1} differs from the actual bound to second order in the perturbations.  Here we give the full expression for the user calculated bound, defined as \cref{eqn:bound1}, where the true values of $\gamma_1$ and $\gamma_2$ have been replaced by their estimated values, and $t$ represents the average gate duration.   
\begin{align}
\epsilon^{\textrm UB} (\text{user}) & \approx 2 \epsilon^{\textrm UB}_{GD} + \frac{t\beta^2}{\Gamma_2' - \Gamma_1} \Big( \frac{1}{2}e^{-\Gamma_2't} -\frac{3}{2}e^{-\Gamma_1t} \Big) + \sqrt{9\sum_{i,j} \mathcal{E}^2_{i,j} (\text{user})} \notag \\
\sum_{i,j} \mathcal{E}^2_{i,j} (\text{user}) & \approx \sum_{i,j} \mathcal{E}_{i,j}^2 + \frac{t\beta^2}{\Gamma_2' - \Gamma_1} \Big( - 6(r_x+r_y)e^{-\Gamma_2't} + 12r_ze^{-\Gamma_1t} + 2e^{-2\Gamma_1t} -6e^{-\Gamma_1t} -2e^{-2\Gamma_2't} +6e^{-\Gamma_2't} \Big)
\end{align}

\section{A robust upper bound on the diamond norm}
Here we add additional terms to the upper bound on the diamond norm of the perturbed generalized damping channel \cref{eqn:bound1}, to obtain $\epsilon^{\textrm UB}_{\text{robust}}$. Terms have been added such that the user calculated version of the new bound $\epsilon^{\textrm UB}_{\text{robust}}$ is greater than the actual bound \cref{eqn:bound1}. Let
\begin{align}
\epsilon^{\textrm UB}_{\text{robust}} &= 2\epsilon^{\textrm UB}_{GD} + 12(r_\text{ideal}^X + r_\text{ideal}^Y - r^X - r^Y) + \sqrt{9\sum \mathcal{E}^2_{\text{robust}}} \geq \epsilon^{\textrm UB} \notag \\
\sum \mathcal{E}^2_{\text{robust}} &= \sum_{i,j} \mathcal{E}^2_{i,j} +6(r_\text{ideal}^X + r_\text{ideal}^Y - r^X - r^Y) \geq \sum_{i,j} \mathcal{E}^2_{i,j}
\end{align} 
where the inequalities follow from $(r_\text{ideal}^X + r_\text{ideal}^Y - r^X - r^Y) \geq 0$ (confirmed numerically where $\beta$ and $\delta$ are allowed to be complex).
As before, we find the user calculated version of the robust bound by replacing all occurrences of $\gamma_1$ and $\gamma_2$ with their estimated values. We obtain
\begin{align}
\epsilon^{\textrm UB}_{\text{robust}}(\text{user}) & \approx [2\epsilon^{\textrm UB}_{GD} + 12(r_\text{ideal}^X + r_\text{ideal}^Y - r^X - r^Y) ]+ \frac{t\beta^2}{\Gamma_2' - \Gamma_1}(\frac{5}{2}e^{-\Gamma_2't} - \frac{3}{2}e^{-\Gamma_1t}) + \sqrt{9\sum \mathcal{E}^2_{\text{robust}} (\text{user})} \\ \notag & \geq \epsilon^{\textrm UB}_{\text{robust}} \geq \epsilon^{\textrm UB} \\ \notag
\sum\mathcal{E}^2_{\text{robust}} (\text{user}) & \approx [\sum_{i,j}\mathcal{E}^2_{i,j} + 6(r_\text{ideal}^X + r_\text{ideal}^Y - r^X - r^Y) ] \\ \notag &+ \frac{t\beta^2}{\Gamma_2' - \Gamma_1}(-6(r^X+r^Y)e^{-\Gamma_2't} + 12r^Ze^{-\Gamma_1t} + 2e^{-2\Gamma_1t} - 6e^{-\Gamma_1t} -2e^{-2\Gamma_2't} +7e^{-\Gamma_2't}) \\ \notag &\geq \sum\mathcal{E}^2_{\text{robust}} \geq \sum_{i,j}\mathcal{E}^2_{i,j} 
\end{align}
where it is helpful to note $r^X,r^Y,r^Z \approx 1/3$.

\section{Numerical comparison of diamond norm bounds}

\begin{figure*}[h]
\centering
\subfigure[~~Scaling of bounds by perturbation strength]{
\includegraphics[scale=0.6]{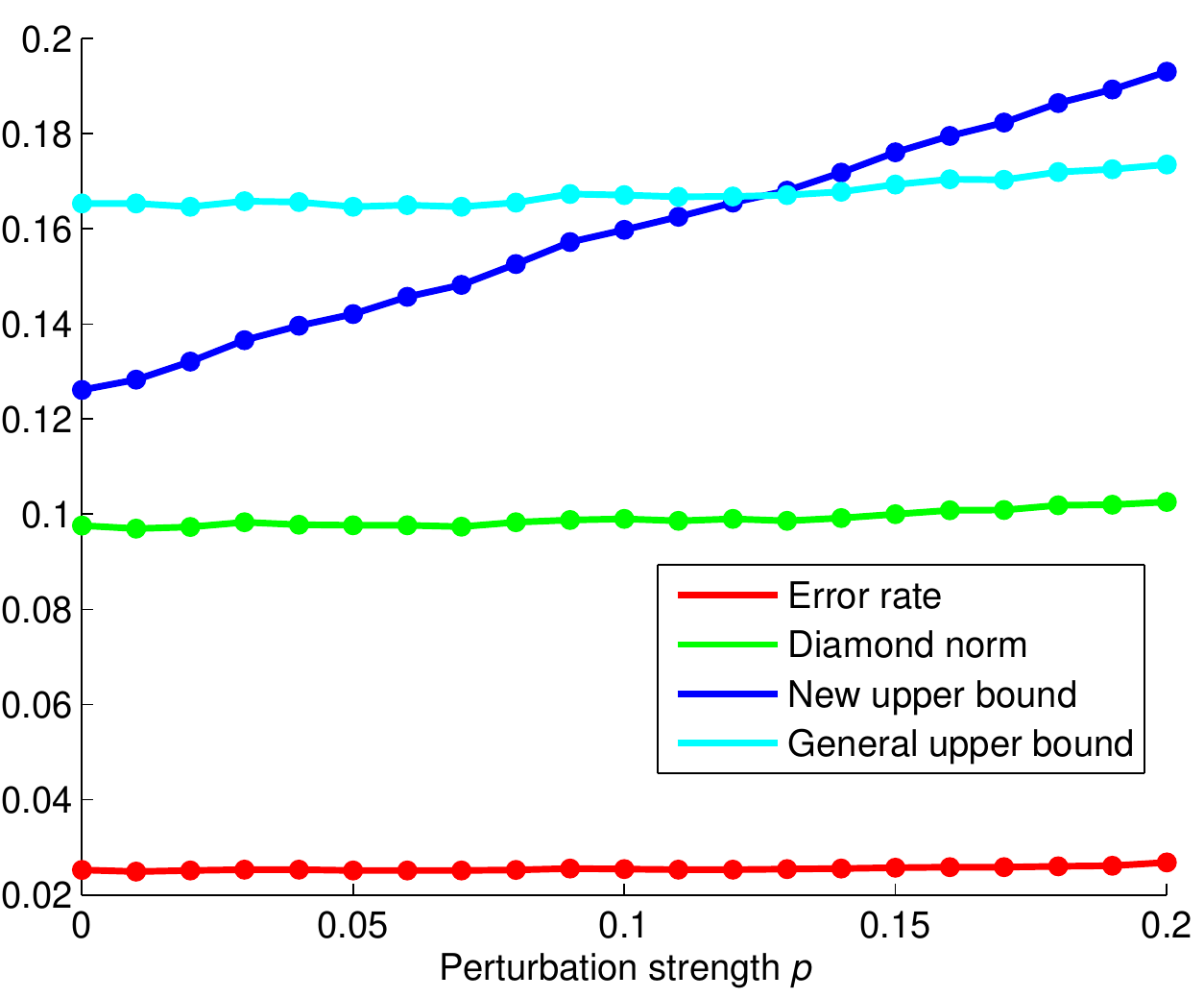}} 
\subfigure[~~Comparison of bounds]{
\includegraphics[scale=0.6]{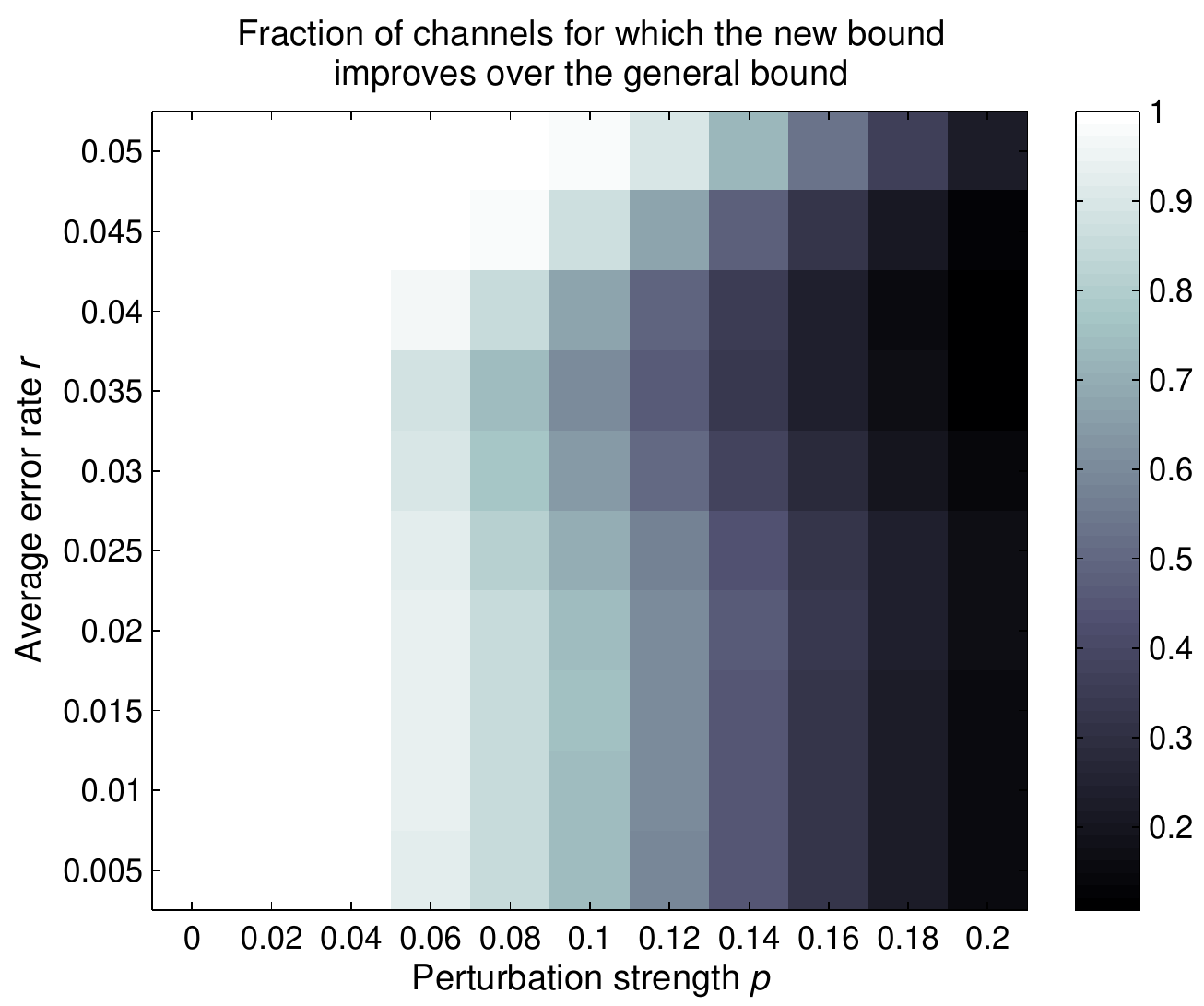}}
\subfigure[~~Tightness of new bound]{
\includegraphics[scale=0.6]{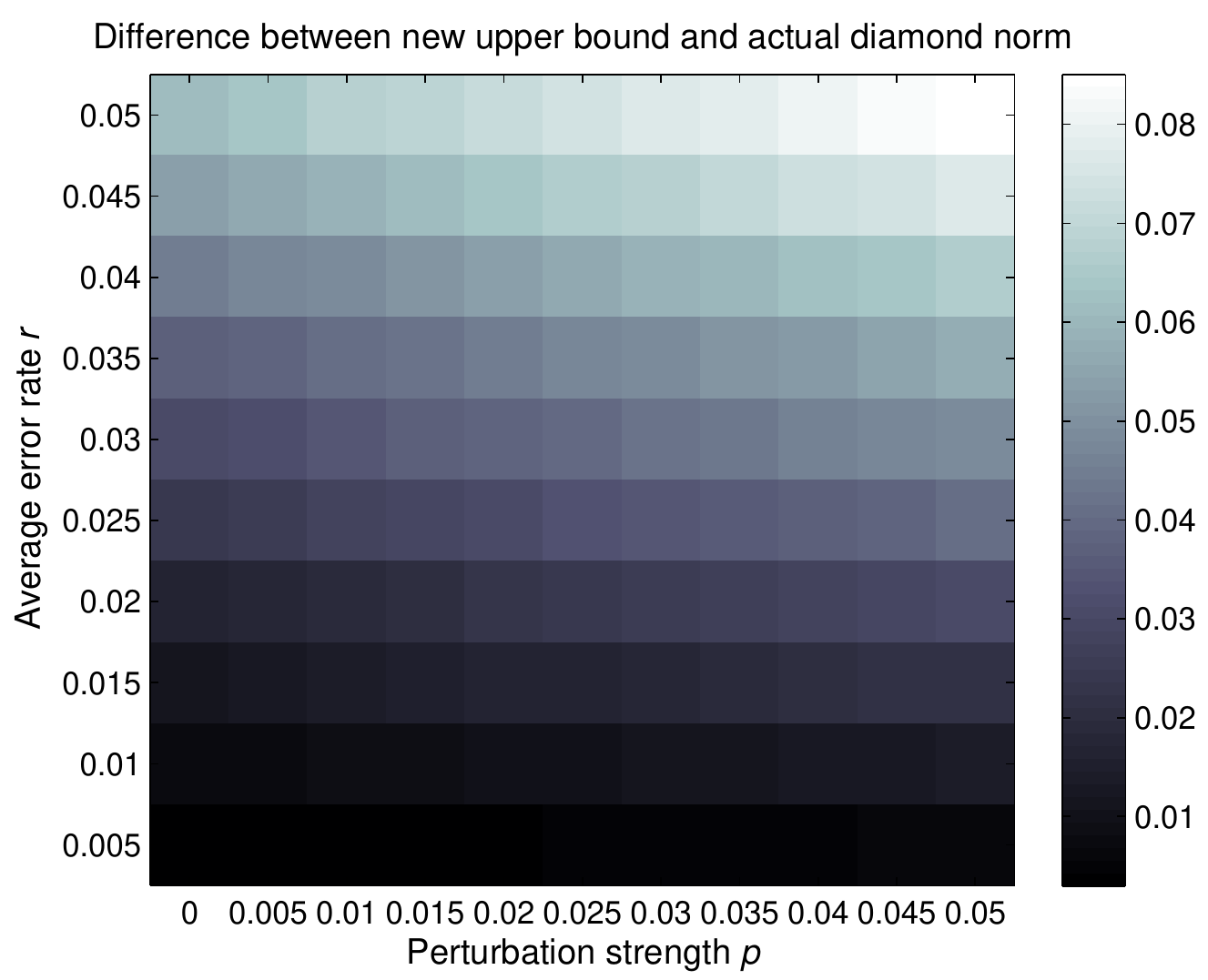}}
\subfigure[~~Improvement over general bound]{
\includegraphics[scale=0.6]{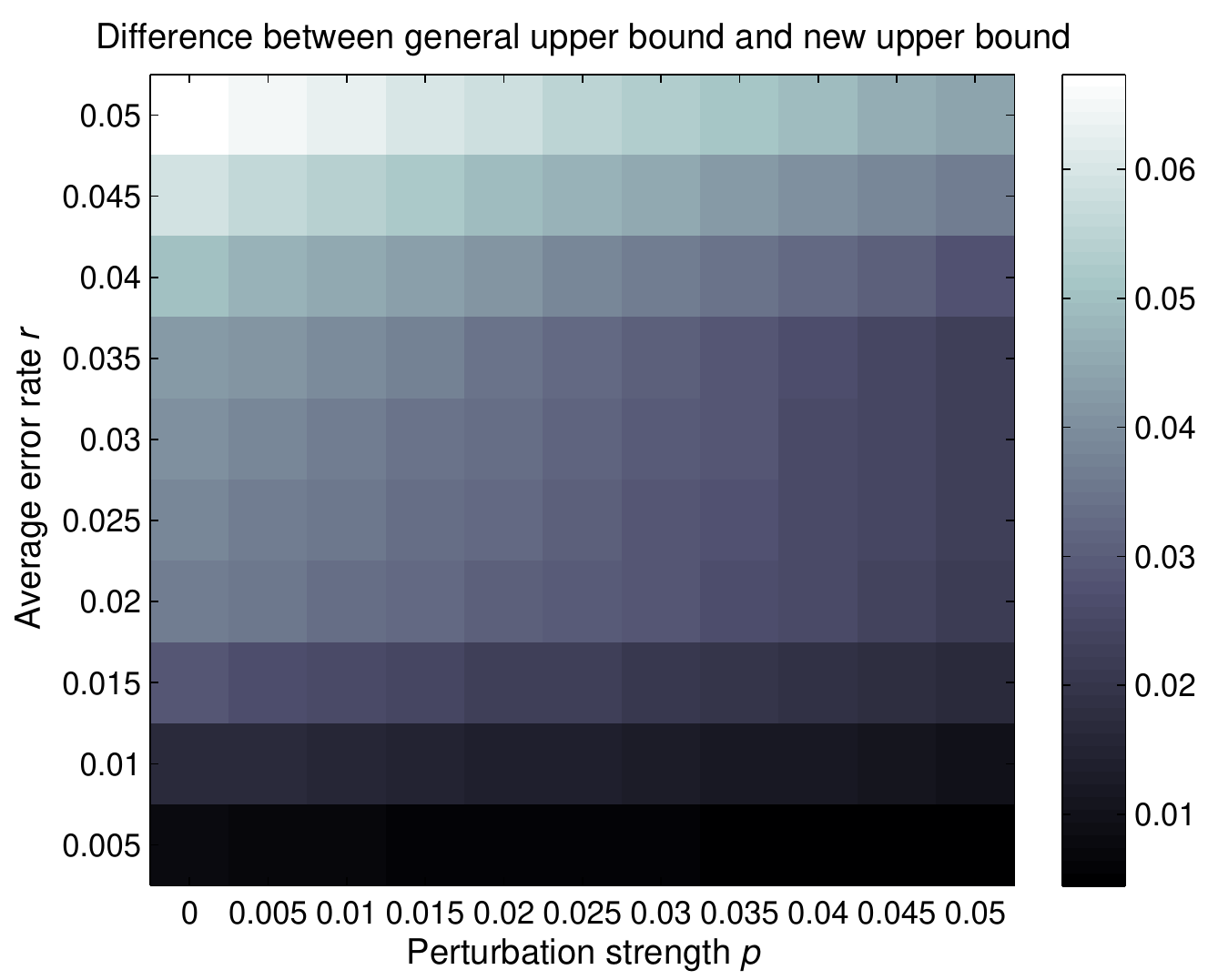}}
\caption{\label{fig:DNnumerics} 
a) The average error rate, diamond norm from the identity channel, and upper bounds on the diamond norm (general given by \cite{Wallman:2015Nov}, new bound given by \cref{eqn:bound1}) averaged over 5000 simulated noise channels where the generalized damping parameters have been set as $\Gamma_2' = 0.1$, $\Gamma_1 = 0.01$, and $\lambda \in [0.8,1]$. The average gate duration has been set to $t = 1 = 10\Gamma_2'$ ($\gamma_1 \approx \Gamma_1$, $\gamma_2 \approx \Gamma_2$). Perturbation parameters $\mathcal{E}_{ij}$ (defined by \cref{eqn:pertChannel}) have been selected uniformly at random in the interval $[-p\Gamma_2',p\Gamma_2']$, which defines the perturbation strength $p$ as the relative size of the perturbation components compared to the leading generalized damping parameter $\Gamma_2'$. Note that this perturbation model allows more freedom than the model described by \cref{eq:general_A} (Hamiltonian terms contributing to a unital perturbation are included). 
b) A heatmap showing the fraction of simulated channels for which \cref{eqn:bound1} obtained a better bound on the diamond norm than the general bound, as a function of the perturbation strength $p$ and average error rate $r$. Each $r$-$p$ bin contains at least 12000 simulated channels. The results show that for perturbation strengths of less than $5\%$, $100\%$ of simulated channels obtained an improved bound, for all values of $r$.   
c-d) The difference between the new upper bound \cref{eqn:bound1} and the actual diamond norm (c) and the difference between the general upper bound and the new upper bound (d) as a function of perturbation strength $p$ and average error rate $r$. Each $r$-$p$ bin is averaged over at least 500 simulated channels. Both the tightness and improvement are on the order of $r$. For example in the smallest $r$-$p$ bin ($p = 0$, $0\leq r \leq .005$) the average diamond norm is $0.0119$, the new upper bound is $0.0148$ (tightness $= 0.0029$), and the general upper bound is $.0223$ (improvement $=0.0075$). 
}
\end{figure*}

Please see \cref{fig:DNnumerics}

\end{document}